\documentclass[apl,aip,twocolumn,reprint,superscriptaddress]{revtex4-1}

\usepackage{bm}
\usepackage[colorlinks=true,linkcolor=blue,citecolor=blue]{hyperref}
\usepackage{times}
\usepackage{amsmath}
\usepackage{amssymb}
\usepackage{amsthm}
\usepackage{amsfonts}
\usepackage{enumerate}
\usepackage{latexsym}
\usepackage{ifpdf}
\newcommand{\beq}{\begin{equation}}
\newcommand{\eeq}{\end{equation}}
\usepackage{graphicx}
\usepackage{makeidx}
\usepackage{color}

\begin{document}

\title{Epitaxial stabilization of ultra thin films of high entropy perovskite}
\author {Ranjan Kumar Patel}
\affiliation  {Department of Physics, Indian Institute of Science, Bengaluru  560012, India}
\author {Shashank Kumar Ojha}
\affiliation  {Department of Physics, Indian Institute of Science, Bengaluru  560012, India}
\author {Siddharth Kumar}
\affiliation  {Department of Physics, Indian Institute of Science, Bengaluru  560012, India}
\author {Akash Saha}
\affiliation  {Undergraduate Programme, Indian Institute of Science, Bengaluru 560012, India}
\author {Prithwijit Mandal}
\affiliation  {Department of Physics, Indian Institute of Science, Bengaluru  560012, India}
\author{J. W. Freeland}
\affiliation {Advanced Photon Source, Argonne National Laboratory, Argonne, Illinois 60439, USA}
\author {S. Middey}
\email{smiddey@iisc.ac.in}
\affiliation  {Department of Physics, Indian Institute of Science, Bengaluru  560012, India}

\begin{abstract}
High entropy oxides (HEOs) are a class of materials, containing equimolar portions of five or more transition metal and/or rare-earth elements. We report here about the layer-by-layer growth of   HEO [(La$_{0.2}$Pr$_{0.2}$Nd$_{0.2}$Sm$_{0.2}$Eu$_{0.2}$)NiO$_3$] thin films on NdGaO$_3$ substrates by pulsed laser deposition. The combined characterizations with in-situ reflection high energy electron diffraction, atomic force microscopy, and X-ray diffraction affirm the single crystalline nature of the film with smooth surface morphology. The desired +3 oxidation of Ni has been confirmed by  an element sensitive X-ray absorption spectroscopy measurement.  Temperature dependent electrical transport measurements revealed a first order metal-insulator transition with the transition temperature very similar to the undoped NdNiO$_3$. Since both of these systems have a comparable tolerance factor, this work demonstrates that the electronic behaviors of $A$-site disordered perovskite-HEOs are primarily controlled by the average tolerance factor.

  \end{abstract}

\maketitle


Finding new materials and new ways to tune material's properties are essential to fulfill the demand of the constantly evolving modern technology. Transition metal oxides show various fascinating electronic and magnetic phenomena such as metal-insulator transition, superconductivity, colossal magnetoresistance, multiferroicity, skyrmions, etc., which have lots of prospect for technological applications~\cite{Imada:1998p1039,Tokura:2006p797,Yang:2011p337,Catalan:2012p119,Lorenz:2016p433001,Matsuno:2018p899}. Furthermore, transition metal (TM) based high entropy oxides (HEOs) are being explored  in recent years to achieve tunable properties in unexplored parts of complex phase diagram~\cite{Rost:2015p8485,Berardan:2016p9536,Rost:2017p2732,Sarkar:2017p747,Djenadic:2017p102,Jiang:2018p116,Anand:2018p119,Sarkar:2018p3400,Sharma:2018p060404,Dkabrowa:2018p32,Witte:2019p034406,Sarkar:2019p1806236,Zhang2019plong,Meisenheimer:2017p13344,Sharma:2018p060404,sharma:2019magnetic}. In general, the configurational entropy of a multi-component solid solution  can be enhanced by mixing a large number of cations in equiatomic proportions and a single structural phase is formed if the entropy contribution overcomes enthalpy driven phase separation ($\Delta G_\mathrm{mix}$=$\Delta H_\mathrm{mix}$-$T\Delta S_\mathrm{mix}$; $\Delta G_\mathrm{mix}$, $\Delta H_\mathrm{mix}$, $\Delta S_\mathrm{mix}$ are Gibbs free energy, enthalpy and entropy of mixing, respectively)~\cite{Rost:2015p8485,Sarkar:2019p1806236}. After the report of the first HEO [Mg$_{0.2}$Ni$_{0.2}$Co$_{0.2}$Cu$_{0.2}$Zn$_{0.2}$O with rocksalt  structure]  by Rost et al.~\cite{Rost:2015p8485},  HEOs with other structural symmetry such as perovskite~\cite{Sharma:2018p060404,Witte:2019p034406}, spinel~\cite{Dkabrowa:2018p32} have been also synthesized. However, this  promising field of HEO is at a very early stage and most of the aspects of HEOs are yet to be explored experimentally. For example, it is still unknown whether the strong disorder or the average tolerance factor ($t_\mathrm{avg}$)  determines the electronic and magnetic behaviors of  perovskite-HEOs.



\begin{figure}
\begin{center}
\vspace{-0pt}
\includegraphics[width=0.48\textwidth] {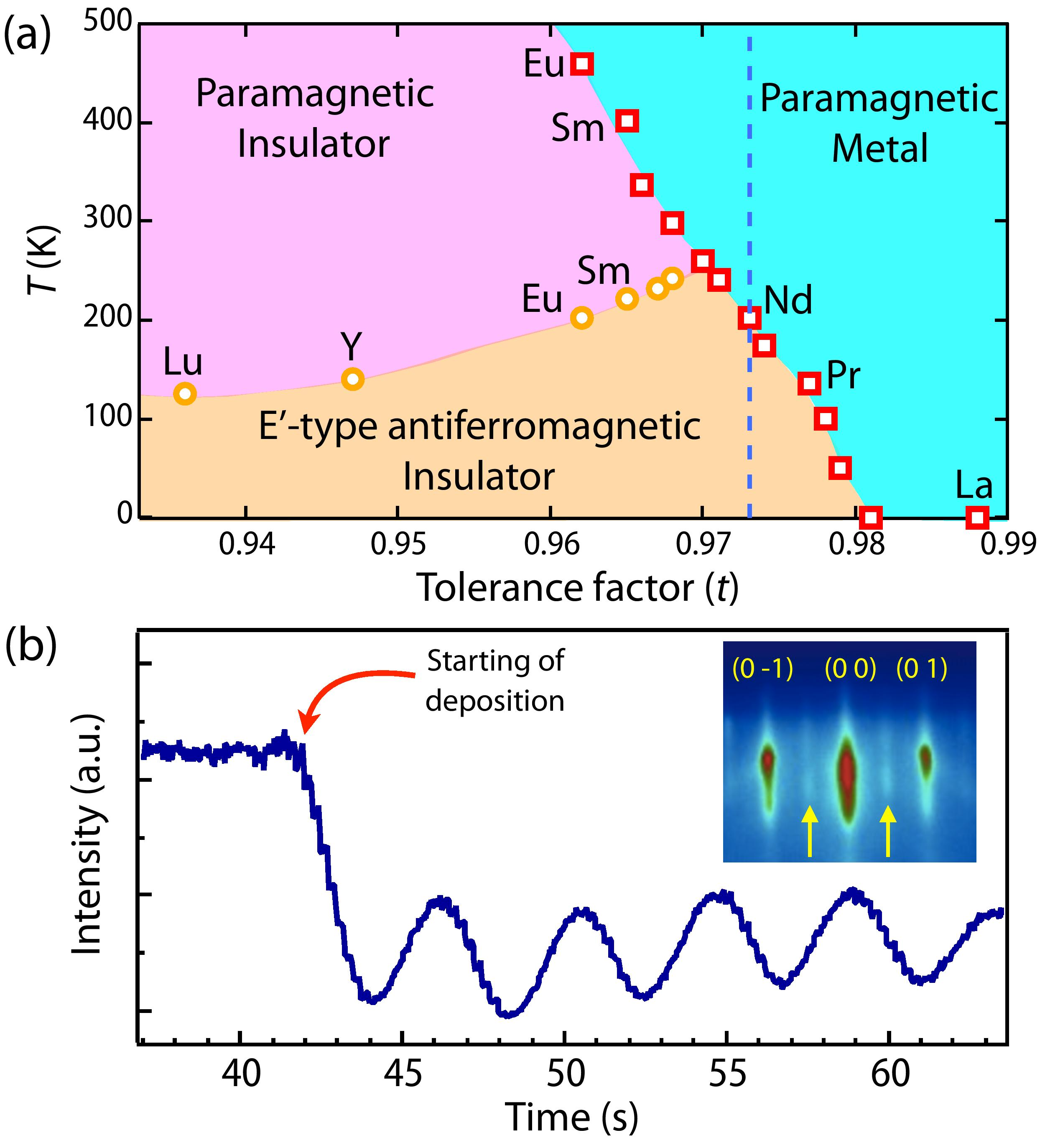}
\caption{\label{Fig1} (a) Phase diagram of $RE$NiO$_3$ series, following Ref.~\onlinecite{Zhou:2003p020404,Meyers:2013p075116}. The dotted line represents the average of the tolerance factors of  LaNiO$_3$, PrNiO$_3$, NdNiO$_3$, SmNiO$_3$, EuNiO$_3$. (b) Temporal variations of specular spot intensity of RHEED pattern during the growth of (LPNSE)NO film  on NGO substrate. RHEED image of the film after cooling is shown in the inset. }
\end{center}
\end{figure}

 As a prototypical example of perovskite ($AB$O$_3$) series, $RE$NiO$_3$  ($RE$= La, Pr, Nd, Sm, Eu...Lu) exhibits an interesting phase diagram as a function of tolerance factor ($t$=$\frac{R_{RE}+R_\mathrm{O}}{\sqrt2{(R_{\mathrm{Ni}}+R_\mathrm{O})}}$, where $R_{RE}$, $R_\mathrm{Ni}$, $R_\mathrm{O}$ are radii of $RE$, Ni and O, respectively)~\cite{Medarde:1997p1679,Catalan:2008p729}.
LaNiO$_3$, the least distorted member of this series remains metallic and paramagnetic down to the lowest temperature. Bulk PrNiO$_3$ and NdNiO$_3$ (NNO) show temperature driven simultaneous transitions from an orthorhombic, paramagnetic, metallic phase to a monoclinic, antiferromagnetic, insulating phase respectively (see Fig.~\ref{Fig1}(a)). The insulating phase is also characterized by a checkerboard type charge ordering~\cite{Staub:2002p126402}. In case of the more distorted members, such as  SmNiO$_3$, EuNiO$_3$, etc, the magnetic transition gets decoupled from the other three simultaneous transitions, resulting in an intermediate paramagnetic, insulating, charge ordered phase. The quest to understand the origin of these transitions have led to remarkable progress in epitaxial stabilization of $RE$NiO$_3$ family (see Refs.~\onlinecite{Middey:2016p305,Catalano:2018p046501}, and literature cited therein), and  thin films  of $RE$NiO$_3$ with $RE$=La, Pr, Nd, Sm, and Eu have been stabilized so far ~\cite{Middey:2016p305,Catalano:2018p046501,Ha:2012p233,Liu:2013p2714,Feigl:2013p51,Meyers:2013p075116,Mikheev:2015p1500797,Hepting:2014p227206,Scherwitzl:2011p246403,Bruno:2013p195108}. This further provides a unique opportunity to verify the role of disorder vs. $t_\mathrm{avg}$, in determining the electronic and magnetic properties of perovskite HEO with a strong disorder at the $A$ site.


\begin{table}[b!]
\caption {Lattice parameters ($a$, $b$, and $c$) of few rare earth nickelates. Pseudo cubic lattice constants $a_{pc}$, $b_{pc}$, and $c_{pc}$ are also listed.}
\begin{tabular}{lcccccc}
\hline
\hline
 Compound    & $a$({\AA})      &  $b$({\AA})   & $c$ ({\AA})  & $a_\mathrm{pc}$= $b_\mathrm{pc}$ ({\AA}) &$c_\mathrm{pc}$ ({\AA})  & Referrence\\ \hline
 LaNiO$_3$ & 5.457 & 5.457 & 13.146 & 3.838 &3.838  &\onlinecite{Garica:1992p4414} \\
 PrNiO$_3$ & 5.419 & 5.380 & 7.626 & 3.818 & 3.813  & \onlinecite{Garica:1992p4414}\\
  NdNiO$_3$ & 5.389 & 5.382 & 7.610 & 3.808 & 3.805   & \onlinecite{Garica:1992p4414}\\
  SmNiO$_3$ & 5.327 & 5.432 & 7.565 &   3.804&3.782   & \onlinecite{Alonso:1999p4754}\\
   EuNiO$_3$ & 5.294 & 5.458 & 7.537 &   3.802& 3.769  & \onlinecite{Alonso:1999p4754} \\\hline
\end{tabular}
\end{table}

In this letter, we report successful layer-by-layer epitaxial growth of  (La$_{0.2}$Pr$_{0.2}$Nd$_{0.2}$Sm$_{0.2}$Eu$_{0.2}$)NiO$_3$ [(LPNSE)NO] thin films on a single crystalline NdGaO$_3$ (NGO) substrate by pulsed laser deposition (PLD). The variation in bulk lattice constants together with the pseudo-cubic lattice constants of several members of the $RE$NiO$_3$ series have been listed in Table-I.
Since the average of the $t$ of $RE$NiO$_3$ with $RE$=La, Pr, Nd, Sm, Eu (indicated by a vertical line in Fig.~\ref{Fig1}(a)) is comparable to NNO, the electronic behavior of [(LPNSE)NO] thin films have been also compared with NNO films. Several characterization techniques including in-situ RHEED (reflection high energy electron diffraction) and ex-situ atomic force microscopy (AFM), X-ray diffraction (XRD), X-ray absorption spectroscopy (XAS) confirmed high structural quality of these (LPNSE)NO] thin films with proper oxidation state of Ni. Transport measurements and XAS experiments further revealed that in-spite of having a strong structural disorder, the electronic behaviors of (LPNSE)NO sample are very similar to a single $A$ site cation NNO film.

(LPNSE)NO films with thickness 15 uc, 30 uc and 45 uc (uc=unit cell in pseudocubic notation) and NNO film with 15 uc were grown on NGO (1 1 0)$_\mathrm{or}$ [(0 0 1)$_\mathrm{pc}$] substrates (here or and pc denote orthorhombic and pseudocubic setting)  by a PLD system  at 735$^\circ$C under a dynamic oxygen pressure of 100-150 millitorr. A KrF excimer laser, operating with 4 Hz and energy density 1.5 J/cm$^2$ was used for the deposition.  The layer by layer growth was monitored by a high pressure RHEED system. The films were post-annealed at the growth temperature under an oxygen pressure of 500 torr  for 30 minutes.   A Park system AFM was used to check the morphology of these films. X-ray diffraction patterns were recorded using a Rigaku Smartlab X-ray diffractometer. Temperature dependent resistivity was measured by using the Van der Pauw geometry in a Quantum Design PPMS (physical property measurement system). XAS spectra at Ni-$L_{3,2}$ and O-$K$ edges were collected in bulk-sensitive TFY (total fluorescence yield) mode  at the 4-ID-C beamline of Advanced Photon Source, Argonne National Laboratory.

The time dependent intensity of the specular reflection of RHEED pattern  (Fig. 1(b)), recorded during the deposition shows very  prominent oscillations, confirming the layer-by-layer growth of (LPNSE)NO film. The inset of Fig. 1(b) shows a RHEED image of a (LPNSE)NO film, taken after cooling to the room temperature. The streaky pattern of specular (0 0) and off-specular (0 1), (0 -1) Bragg reflections is a characteristic of smooth surface morphology. Akin  to the RHEED pattern of NNO film on NGO substrate~\cite{Ojha:2019p235153}, (LPNSE)NO films also have half-order spots: (0 1/2) and (0 -1/2) (denoted by the arrows),  implying orthorhombic/monoclinic symmetry at room temperature.   Inset of Fig. 2(a) shows AFM image of the   45 uc (LPNSE)NO film and  the  roughness  is found to be $\sim$1.8\AA\, well below the $c_\mathrm{pc}$, further testifying excellent surface morphology of the film.

\begin{figure}
\begin{center}
\vspace{-0pt}
\includegraphics[width=0.48\textwidth] {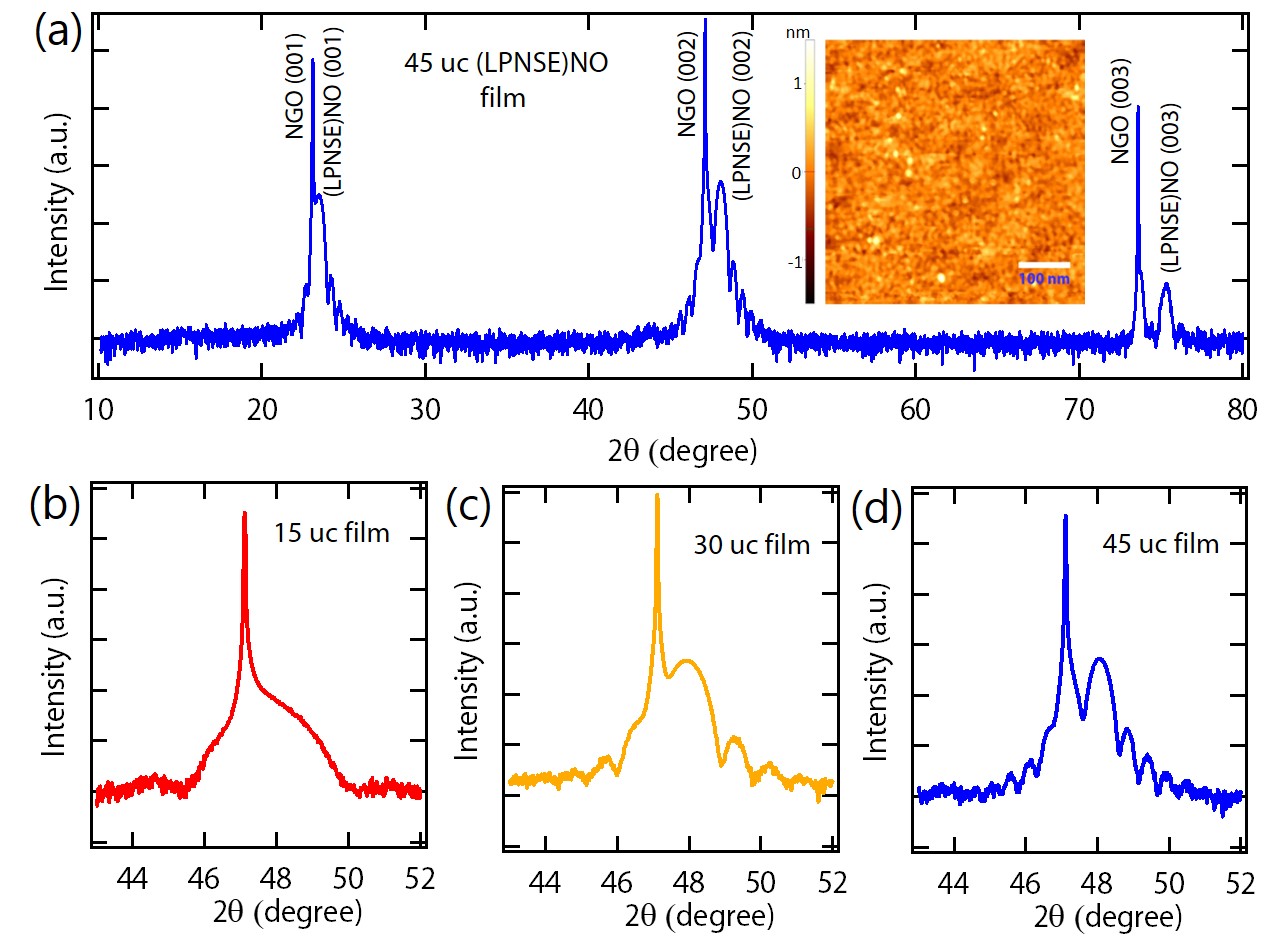}
\caption{\label{Fig2} (a) Long 2$\theta$-$\omega$ XRD scan for 45 uc (LPNSE)NO film.  (b)-(d) XRD patterns of 15, 30, and 45 uc films near NGO (0 0 2)$\mathrm{pc}$ peak, respectively.  AFM surface morphology of the 45 uc film is shown in the inset of (a).}
\end{center}
\end{figure}

In order to check the structural quality of the samples and to detect the presence of any impurity phase, we have recorded 2$\theta$-$\omega$ diffraction scan for (LPNSE)NO films  using Cu $K_\alpha$ radiation. Such a long scan XRD for the 45 uc (LPNSE)NO film (Fig. ~\ref{Fig2}(a)) consists of broad film peaks in the vicinity of sharp substrate peaks, confirming the single crystalline nature of the film. Most importantly, the absence  of any impurity peaks (within the detection limit of XRD)  infers  the growth conditions, used in this work is able to stabilize the multicomponent system into a single phase. XRD patterns around the (0 0 2)$_\mathrm{pc}$ substrate peak for 15 uc, 30 uc, and 45 uc (LPNSE)NO films are shown in Fig. ~\ref{Fig2}(b), (c), and (d), respectively. The very close proximity between the film peak and  the substrate peak (Fig. ~\ref{Fig2}(b)) in  case of the 15 uc LPNSE)NO film prohibits  a reliable estimation  of out-of-plane lattice constant ($c_\mathrm{pc}$). $c_\mathrm{pc}$ for 30 uc and 45 uc film are found to be 3.792\AA\ and   3.784\AA. The presence of thickness fringes in the vicinity of the film peaks further supports the excellent flatness of the film-substrate interface. The thickness of the films calculated from the position of the fringes (e.g. $\sim$17.3 nm for 45 uc film) are very close to the value expected from the RHEED oscillations.  $c_\mathrm{pc}$ of the 15 uc NNO film is found to be 3.845\AA\ (XRD pattern not shown).

\begin{figure}
\begin{center}
\vspace{-0pt}
\includegraphics[width=0.49\textwidth] {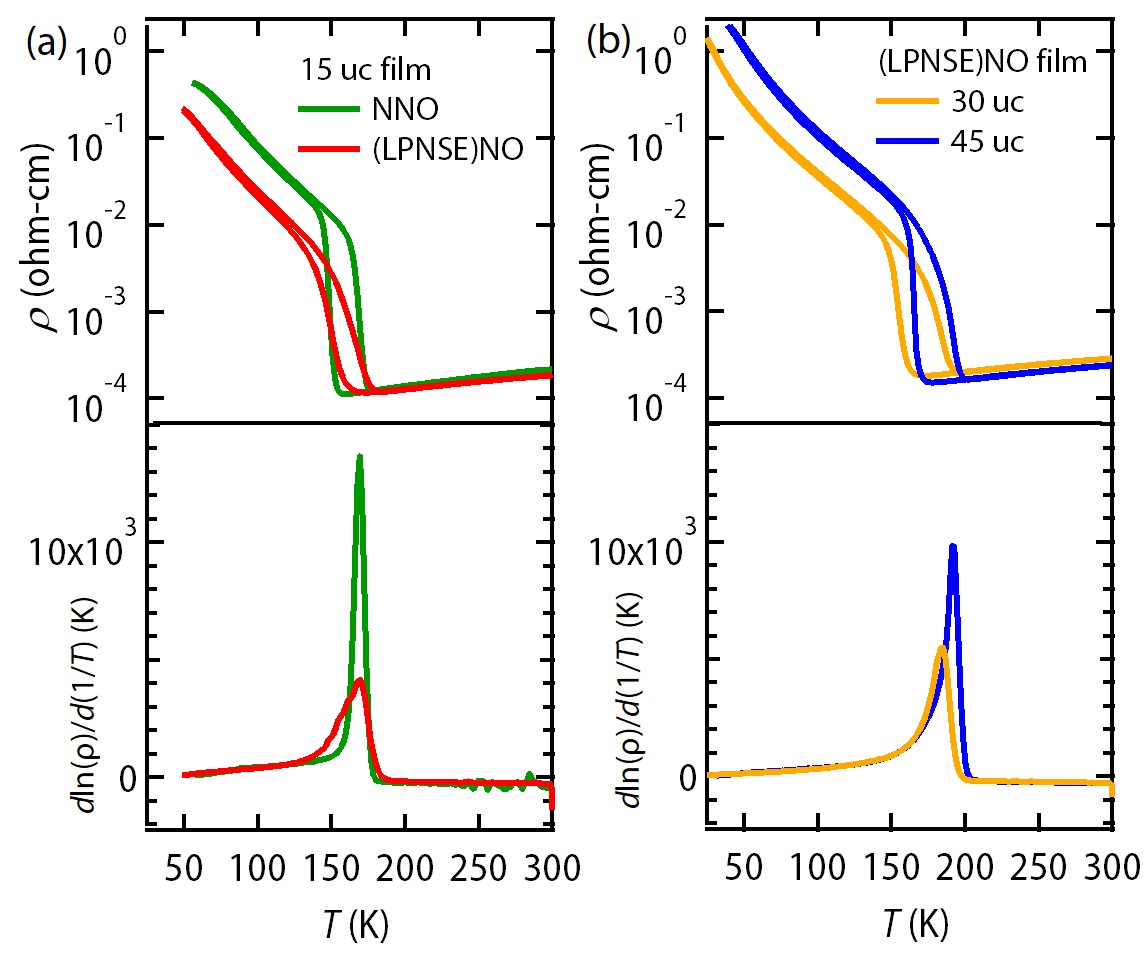}
\caption{\label{transport}   Temperature dependent resistivity of 15 uc NNO and 15 uc (LPNSE)NO film [30 uc and 45 uc (LPNSE)NO film] are shown in the upper panel of (a) [(b)].  Estimation of $T_N$ from $d$ln$(\rho)$/$d{(1/T)}$ vs. $T$ analysis are plotted in the corresponding lower panel.}
\end{center}
\end{figure}

After confirming the high structural and morphological quality, we have investigated the electrical transport of the films. As reported earlier~\cite{Liu:2013p2714,Mikheev:2015p1500797,Ojha:2019p235153},  15 uc NNO thin film on NGO substrate undergoes first order MIT (upper panel of Fig.~\ref{transport}(a)). The transition temperature in the cooling run ($T^c_\mathrm{MIT}\sim$ 160 K) and heating ($T^h_\mathrm{MIT}\sim$ 180 K) is lower compared to the bulk NNO and is related to the epitaxial strain and finite thickness~\cite{Liu:2013p2714}. Surprisingly, the resistivity ($\rho$) of 15 uc (LPNSE)NO film at 300 K is very similar to that of 15 uc NNO film in-spite of having a strong disorder on the $A$ site. It further exhibits a MIT  with strong thermal hysteresis ($T^c_\mathrm{MIT}\sim$ 175 K, $T^h_\mathrm{MIT}\sim$ 185 K). However, the transition is more sluggish and $\rho$ of the insulating phase is also lower than that of 15 uc NNO film. With the increase of the film thickness, $T_\mathrm{MIT}$ becomes approximately 200 K ((upper panel of Fig.~\ref{transport}(b)), which is very close to the transition temperature expected for the corresponding  $t$ of (LPNSE)NO phase from the bulk phase diagram (Fig. 1(a)). This finding clearly establishes that the average tolerance factor  controls the $T_\mathrm{MIT}$ for this HEO,  rather than the disorder at $A$-site.

 All $RE$NiO$_3$ with an insulating phase also host $E^\prime$-type antiferromagnetic ordering~\cite{Garcia:1994p978,Scagnoli:2006p100409,Liu:2013p2714,Hepting:2014p227206,Middey:2018p081602,Middey:2018p156801}. The magnetic transition temperature ($T_N$) can be approximately estimated from $d$(ln$\rho$)/$d$(1/$T$) vs. $T$ plot, as demonstrated recently for NNO films and EuNiO$_3$/LaNiO$_3$ superlattices~\cite{Middey:2018p156801,Middey:2018ENOLNOStrain,Ojha:2019p235153}. Such resistivity analysis of the heating run data (lower panel of Fig.~\ref{transport}(a), (b)) provides a $T_N$ of 170 K for both 15 uc NNO and 15 uc (LPNSE)NO sample, 185 and 190 K for 30 uc and 45 uc (LPNSE)NO film, respectively.  This suggests a simultaneous electronic and magnetic transitions in these (LPNSE)NO films. Soft X-ray resonant scattering experiments can further confirm this~\cite{Scagnoli:2006p100409,Liu:2013p2714,Hepting:2014p227206,Middey:2018p081602,Middey:2018p156801}.

\begin{figure}
\begin{center}
\vspace{-0pt}
\includegraphics[width=0.45\textwidth] {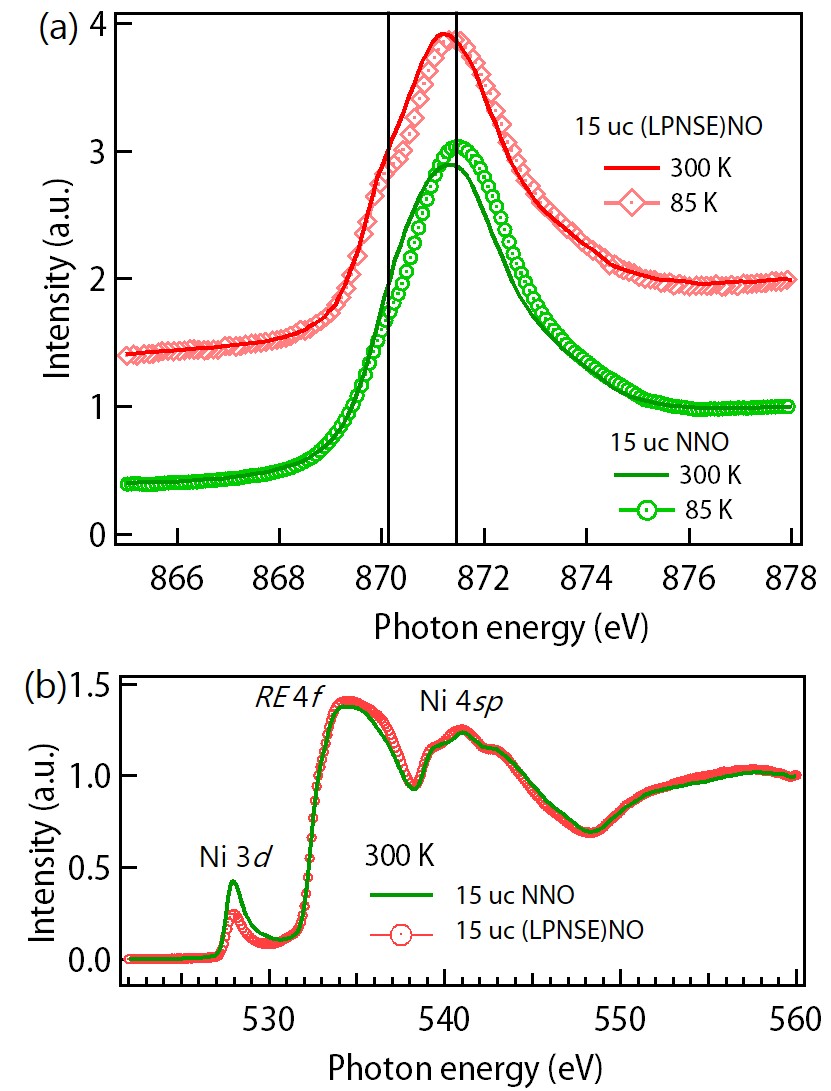}
\caption{\label{xas} (a) Ni $L_2$-edge, (b) O $K$-edge XAS for 6 nm NNO and 6 nm (LPNSE)NO films on NGO substrate. In the O-$K$ edge XAS spectra, the pre-peak around 528 eV  is related to the transition from O $1s$ to the Ni
3$d$ -O 2$p$ hybridized states. 4$f$ and 5$d$ hybridized states of rare earth ($RE$) ions, present in sample and substrate results the peak around 534-535 eV. Ni 4$s$ and 4$p$ also hybridized with O 2$p$, resulting the peaks around 540-545 eV.}
\end{center}
\end{figure}

The required  high +3 oxidation state of Ni  makes $RE$NiO$_3$ based systems very susceptible to the oxygen nonstoichiometry. In order to further understand the electronic and chemical structure of these films, we have measured XAS spectra on Ni $L_{3,2}$ edge and O-$K$ edge at 300 K and 85 K, i.e. much below $T_\mathrm{MIT}$. Due to strong overlap of La $M_4$ edge with Ni $L_3$ for (LPNSE)NO film, we discuss here only $L_2$ edge. First of all, the XAS line shape of both 15 uc NNO and 15 uc (LPNSE)NO film  at 300 K (Fig.~\ref{xas}(a)) is consistent with Ni$^{3+}$ in the metallic phase of nickelates~\cite{Medarde:1992p14975,Liu:2010p233110,Wu:2015p195130,Middey:2014p6819},  affirming the stabilization of desired oxidation state of Ni. Further, the appearance of strong multiple structures (around 870.05 eV) in the insulating phase of the (LPNSE)NO film  is also consistent with the observation of NNO film (Fig.~\ref{xas}(a)) and the insulating phase of other nickelates~\cite{Medarde:1992p14975,Liu:2010p233110,Meyers:2013p075116,Wu:2015p195130}. Similar to the high $T_c$ cuprates, $RE$NiO$_3$ also contains ligand holes, which can be observed as a pre-peak around 528 eV in O $K$ edge XAS spectrum due to the $d^8\underline{L}\rightarrow\underline{c}d^8$ ( $\underline{c}$ denotes hole in the oxygen 1$s$ core state and $\underline{L}$ corresponds to a hole in O 2$p$ state)~\cite{Medarde:1992p14975,Liu:2013p2714,Middey:2014p6819,Middey:2018ENOLNOStrain,Freeland:2016p56}. While the  intensity of the pre peak is  reduced in (LPNSE)NO film  at 300 K,  the  position and FWHM (full width at half maxima) are very similar for both samples. The measurements at 85 K have found lowering of peak width in both samples (not shown), which is expected due to the band narrowing across the MIT~\cite{Meyers:2016p27934,Freeland:2016p56}. Thus,  transport and XAS measurements conclude that the overall electronic structure effect across the MIT of (LPNSE)NO film is very similar to that of   NNO film.

To summarize, we have successfully grown high quality epitaxial films of multicomponent (LPNSE)NO in a layer-by-layer fashion by pulsed laser deposition. RHEED, XRD, AFM, XAS and transport measurements  have been carried out to investigate the structure and the electronic behavior of these films. In spite of having multi elements and strong disorder at the $RE$ site, the average tolerance factor determines the electronic transition temperature. However, the microscopic details e.g. nucleation of insulating/metallic phase around the transition temperature~\cite{post:2018p1056}, charge transfer between Ni and $RE$ sites~\cite{Upton:2015p036401}, conductivity noise~\cite{Daptary:2019p125105}, etc. may depend on the on the details of $RE$ composition  and needs to be explored further. Stabilization of such multi $RE$ system in a single crystalline form would further allow to investigate the complexity of phase transitions around the triple point of $RE$NiO$_3$ phase diagram~\cite{Park:2013p431}.   Due to the chemical diversity of the perovskite family,  huge numbers of HEO with a strong disorder at either $A$ site or $B$ site or both sites can be studied in the future to explore disorder driven physics in  strongly correlated systems.

This work is funded by  a DST Nanomission grant (DST/NM/NS/2018/246) and a SERB Early Career Research Award (ECR/2018/001512). The authors acknowledge AFM and XRD facilities at the Department
of Physics, IISc Bangalore. RKP and SM thank the Department of Science and Technology, India (SR/NM/Z-07/2015) for the financial support to conduct synchrotron experiment at Advanced Photon Source and Jawaharlal Nehru Centre for Advanced Scientific Research (JNCASR) for managing the project. This research used resources of the Advanced Photon Source, a U.S. Department of Energy Office of Science User Facility operated by Argonne National Laboratory under Contract No. DE-AC02-06CH11357.


\begin{thebibliography}{53}%
\makeatletter
\providecommand \@ifxundefined [1]{%
 \@ifx{#1\undefined}
}%
\providecommand \@ifnum [1]{%
 \ifnum #1\expandafter \@firstoftwo
 \else \expandafter \@secondoftwo
 \fi
}%
\providecommand \@ifx [1]{%
 \ifx #1\expandafter \@firstoftwo
 \else \expandafter \@secondoftwo
 \fi
}%
\providecommand \natexlab [1]{#1}%
\providecommand \enquote  [1]{``#1''}%
\providecommand \bibnamefont  [1]{#1}%
\providecommand \bibfnamefont [1]{#1}%
\providecommand \citenamefont [1]{#1}%
\providecommand \href@noop [0]{\@secondoftwo}%
\providecommand \href [0]{\begingroup \@sanitize@url \@href}%
\providecommand \@href[1]{\@@startlink{#1}\@@href}%
\providecommand \@@href[1]{\endgroup#1\@@endlink}%
\providecommand \@sanitize@url [0]{\catcode `\\12\catcode `\$12\catcode
  `\&12\catcode `\#12\catcode `\^12\catcode `\_12\catcode `\%12\relax}%
\providecommand \@@startlink[1]{}%
\providecommand \@@endlink[0]{}%
\providecommand \url  [0]{\begingroup\@sanitize@url \@url }%
\providecommand \@url [1]{\endgroup\@href {#1}{\urlprefix }}%
\providecommand \urlprefix  [0]{URL }%
\providecommand \Eprint [0]{\href }%
\providecommand \doibase [0]{http://dx.doi.org/}%
\providecommand \selectlanguage [0]{\@gobble}%
\providecommand \bibinfo  [0]{\@secondoftwo}%
\providecommand \bibfield  [0]{\@secondoftwo}%
\providecommand \translation [1]{[#1]}%
\providecommand \BibitemOpen [0]{}%
\providecommand \bibitemStop [0]{}%
\providecommand \bibitemNoStop [0]{.\EOS\space}%
\providecommand \EOS [0]{\spacefactor3000\relax}%
\providecommand \BibitemShut  [1]{\csname bibitem#1\endcsname}%
\let\auto@bib@innerbib\@empty
\bibitem [{\citenamefont {Imada}, \citenamefont {Fujimori},\ and\ \citenamefont
  {Tokura}(1998)}]{Imada:1998p1039}%
  \BibitemOpen
  \bibfield  {author} {\bibinfo {author} {\bibfnamefont {M.}~\bibnamefont
  {Imada}}, \bibinfo {author} {\bibfnamefont {A.}~\bibnamefont {Fujimori}}, \
  and\ \bibinfo {author} {\bibfnamefont {Y.}~\bibnamefont {Tokura}},\ }\href
  {\doibase 10.1103/RevModPhys.70.1039} {\bibfield  {journal} {\bibinfo
  {journal} {Rev. Mod. Phys.}\ }\textbf {\bibinfo {volume} {70}},\ \bibinfo
  {pages} {1039} (\bibinfo {year} {1998})}\BibitemShut {NoStop}%
\bibitem [{\citenamefont {Tokura}(2006)}]{Tokura:2006p797}%
  \BibitemOpen
  \bibfield  {author} {\bibinfo {author} {\bibfnamefont {Y.}~\bibnamefont
  {Tokura}},\ }\href {\doibase 10.1088/0034-4885/69/3/r06} {\bibfield
  {journal} {\bibinfo  {journal} {Reports on Progress in Physics}\ }\textbf
  {\bibinfo {volume} {69}},\ \bibinfo {pages} {797} (\bibinfo {year}
  {2006})}\BibitemShut {NoStop}%
\bibitem [{\citenamefont {Yang}, \citenamefont {Ko},\ and\ \citenamefont
  {Ramanathan}(2011)}]{Yang:2011p337}%
  \BibitemOpen
  \bibfield  {author} {\bibinfo {author} {\bibfnamefont {Z.}~\bibnamefont
  {Yang}}, \bibinfo {author} {\bibfnamefont {C.}~\bibnamefont {Ko}}, \ and\
  \bibinfo {author} {\bibfnamefont {S.}~\bibnamefont {Ramanathan}},\
  }\href@noop {} {\bibfield  {journal} {\bibinfo  {journal} {Annual Review of
  Materials Research}\ }\textbf {\bibinfo {volume} {41}},\ \bibinfo {pages}
  {337} (\bibinfo {year} {2011})}\BibitemShut {NoStop}%
\bibitem [{\citenamefont {Catalan}\ \emph {et~al.}(2012)\citenamefont
  {Catalan}, \citenamefont {Seidel}, \citenamefont {Ramesh},\ and\
  \citenamefont {Scott}}]{Catalan:2012p119}%
  \BibitemOpen
  \bibfield  {author} {\bibinfo {author} {\bibfnamefont {G.}~\bibnamefont
  {Catalan}}, \bibinfo {author} {\bibfnamefont {J.}~\bibnamefont {Seidel}},
  \bibinfo {author} {\bibfnamefont {R.}~\bibnamefont {Ramesh}}, \ and\ \bibinfo
  {author} {\bibfnamefont {J.~F.}\ \bibnamefont {Scott}},\ }\href@noop {}
  {\bibfield  {journal} {\bibinfo  {journal} {Reviews of Modern Physics}\
  }\textbf {\bibinfo {volume} {84}},\ \bibinfo {pages} {119} (\bibinfo {year}
  {2012})}\BibitemShut {NoStop}%
\bibitem [{\citenamefont {Lorenz}\ \emph {et~al.}(2016)\citenamefont {Lorenz},
  \citenamefont {Rao}, \citenamefont {Venkatesan}, \citenamefont {Fortunato},
  \citenamefont {Barquinha}, \citenamefont {Branquinho}, \citenamefont
  {Salgueiro}, \citenamefont {Martins}, \citenamefont {Carlos}, \citenamefont
  {Liu} \emph {et~al.}}]{Lorenz:2016p433001}%
  \BibitemOpen
  \bibfield  {author} {\bibinfo {author} {\bibfnamefont {M.}~\bibnamefont
  {Lorenz}}, \bibinfo {author} {\bibfnamefont {M.~R.}\ \bibnamefont {Rao}},
  \bibinfo {author} {\bibfnamefont {T.}~\bibnamefont {Venkatesan}}, \bibinfo
  {author} {\bibfnamefont {E.}~\bibnamefont {Fortunato}}, \bibinfo {author}
  {\bibfnamefont {P.}~\bibnamefont {Barquinha}}, \bibinfo {author}
  {\bibfnamefont {R.}~\bibnamefont {Branquinho}}, \bibinfo {author}
  {\bibfnamefont {D.}~\bibnamefont {Salgueiro}}, \bibinfo {author}
  {\bibfnamefont {R.}~\bibnamefont {Martins}}, \bibinfo {author} {\bibfnamefont
  {E.}~\bibnamefont {Carlos}}, \bibinfo {author} {\bibfnamefont
  {A.}~\bibnamefont {Liu}},  \emph {et~al.},\ }\href@noop {} {\bibfield
  {journal} {\bibinfo  {journal} {Journal of Physics D: Applied Physics}\
  }\textbf {\bibinfo {volume} {49}},\ \bibinfo {pages} {433001} (\bibinfo
  {year} {2016})}\BibitemShut {NoStop}%
\bibitem [{\citenamefont {Matsuno}\ \emph {et~al.}(2018)\citenamefont
  {Matsuno}, \citenamefont {Fujioka}, \citenamefont {Okuda}, \citenamefont
  {Ueno}, \citenamefont {Mizokawa},\ and\ \citenamefont
  {Katsufuji}}]{Matsuno:2018p899}%
  \BibitemOpen
  \bibfield  {author} {\bibinfo {author} {\bibfnamefont {J.}~\bibnamefont
  {Matsuno}}, \bibinfo {author} {\bibfnamefont {J.}~\bibnamefont {Fujioka}},
  \bibinfo {author} {\bibfnamefont {T.}~\bibnamefont {Okuda}}, \bibinfo
  {author} {\bibfnamefont {K.}~\bibnamefont {Ueno}}, \bibinfo {author}
  {\bibfnamefont {T.}~\bibnamefont {Mizokawa}}, \ and\ \bibinfo {author}
  {\bibfnamefont {T.}~\bibnamefont {Katsufuji}},\ }\href@noop {} {\bibfield
  {journal} {\bibinfo  {journal} {Science and technology of advanced
  materials}\ }\textbf {\bibinfo {volume} {19}},\ \bibinfo {pages} {899}
  (\bibinfo {year} {2018})}\BibitemShut {NoStop}%
\bibitem [{\citenamefont {Rost}\ \emph {et~al.}(2015)\citenamefont {Rost},
  \citenamefont {Sachet}, \citenamefont {Borman}, \citenamefont {Moballegh},
  \citenamefont {Dickey}, \citenamefont {Hou}, \citenamefont {Jones},
  \citenamefont {Curtarolo},\ and\ \citenamefont {Maria}}]{Rost:2015p8485}%
  \BibitemOpen
  \bibfield  {author} {\bibinfo {author} {\bibfnamefont {C.~M.}\ \bibnamefont
  {Rost}}, \bibinfo {author} {\bibfnamefont {E.}~\bibnamefont {Sachet}},
  \bibinfo {author} {\bibfnamefont {T.}~\bibnamefont {Borman}}, \bibinfo
  {author} {\bibfnamefont {A.}~\bibnamefont {Moballegh}}, \bibinfo {author}
  {\bibfnamefont {E.~C.}\ \bibnamefont {Dickey}}, \bibinfo {author}
  {\bibfnamefont {D.}~\bibnamefont {Hou}}, \bibinfo {author} {\bibfnamefont
  {J.~L.}\ \bibnamefont {Jones}}, \bibinfo {author} {\bibfnamefont
  {S.}~\bibnamefont {Curtarolo}}, \ and\ \bibinfo {author} {\bibfnamefont
  {J.-P.}\ \bibnamefont {Maria}},\ }\href@noop {} {\bibfield  {journal}
  {\bibinfo  {journal} {Nature communications}\ }\textbf {\bibinfo {volume}
  {6}},\ \bibinfo {pages} {8485} (\bibinfo {year} {2015})}\BibitemShut
  {NoStop}%
\bibitem [{\citenamefont {B{\'e}rardan}\ \emph {et~al.}(2016)\citenamefont
  {B{\'e}rardan}, \citenamefont {Franger}, \citenamefont {Meena},\ and\
  \citenamefont {Dragoe}}]{Berardan:2016p9536}%
  \BibitemOpen
  \bibfield  {author} {\bibinfo {author} {\bibfnamefont {D.}~\bibnamefont
  {B{\'e}rardan}}, \bibinfo {author} {\bibfnamefont {S.}~\bibnamefont
  {Franger}}, \bibinfo {author} {\bibfnamefont {A.}~\bibnamefont {Meena}}, \
  and\ \bibinfo {author} {\bibfnamefont {N.}~\bibnamefont {Dragoe}},\
  }\href@noop {} {\bibfield  {journal} {\bibinfo  {journal} {Journal of
  Materials Chemistry A}\ }\textbf {\bibinfo {volume} {4}},\ \bibinfo {pages}
  {9536} (\bibinfo {year} {2016})}\BibitemShut {NoStop}%
\bibitem [{\citenamefont {Rost}\ \emph {et~al.}(2017)\citenamefont {Rost},
  \citenamefont {Rak}, \citenamefont {Brenner},\ and\ \citenamefont
  {Maria}}]{Rost:2017p2732}%
  \BibitemOpen
  \bibfield  {author} {\bibinfo {author} {\bibfnamefont {C.~M.}\ \bibnamefont
  {Rost}}, \bibinfo {author} {\bibfnamefont {Z.}~\bibnamefont {Rak}}, \bibinfo
  {author} {\bibfnamefont {D.~W.}\ \bibnamefont {Brenner}}, \ and\ \bibinfo
  {author} {\bibfnamefont {J.-P.}\ \bibnamefont {Maria}},\ }\href@noop {}
  {\bibfield  {journal} {\bibinfo  {journal} {Journal of the American Ceramic
  Society}\ }\textbf {\bibinfo {volume} {100}},\ \bibinfo {pages} {2732}
  (\bibinfo {year} {2017})}\BibitemShut {NoStop}%
\bibitem [{\citenamefont {Sarkar}\ \emph {et~al.}(2017)\citenamefont {Sarkar},
  \citenamefont {Djenadic}, \citenamefont {Usharani}, \citenamefont {Sanghvi},
  \citenamefont {Chakravadhanula}, \citenamefont {Gandhi}, \citenamefont
  {Hahn},\ and\ \citenamefont {Bhattacharya}}]{Sarkar:2017p747}%
  \BibitemOpen
  \bibfield  {author} {\bibinfo {author} {\bibfnamefont {A.}~\bibnamefont
  {Sarkar}}, \bibinfo {author} {\bibfnamefont {R.}~\bibnamefont {Djenadic}},
  \bibinfo {author} {\bibfnamefont {N.~J.}\ \bibnamefont {Usharani}}, \bibinfo
  {author} {\bibfnamefont {K.~P.}\ \bibnamefont {Sanghvi}}, \bibinfo {author}
  {\bibfnamefont {V.~S.}\ \bibnamefont {Chakravadhanula}}, \bibinfo {author}
  {\bibfnamefont {A.~S.}\ \bibnamefont {Gandhi}}, \bibinfo {author}
  {\bibfnamefont {H.}~\bibnamefont {Hahn}}, \ and\ \bibinfo {author}
  {\bibfnamefont {S.~S.}\ \bibnamefont {Bhattacharya}},\ }\href@noop {}
  {\bibfield  {journal} {\bibinfo  {journal} {Journal of the European Ceramic
  Society}\ }\textbf {\bibinfo {volume} {37}},\ \bibinfo {pages} {747}
  (\bibinfo {year} {2017})}\BibitemShut {NoStop}%
\bibitem [{\citenamefont {Djenadic}\ \emph {et~al.}(2017)\citenamefont
  {Djenadic}, \citenamefont {Sarkar}, \citenamefont {Clemens}, \citenamefont
  {Loho}, \citenamefont {Botros}, \citenamefont {Chakravadhanula},
  \citenamefont {K{\"u}bel}, \citenamefont {Bhattacharya}, \citenamefont
  {Gandhi},\ and\ \citenamefont {Hahn}}]{Djenadic:2017p102}%
  \BibitemOpen
  \bibfield  {author} {\bibinfo {author} {\bibfnamefont {R.}~\bibnamefont
  {Djenadic}}, \bibinfo {author} {\bibfnamefont {A.}~\bibnamefont {Sarkar}},
  \bibinfo {author} {\bibfnamefont {O.}~\bibnamefont {Clemens}}, \bibinfo
  {author} {\bibfnamefont {C.}~\bibnamefont {Loho}}, \bibinfo {author}
  {\bibfnamefont {M.}~\bibnamefont {Botros}}, \bibinfo {author} {\bibfnamefont
  {V.~S.}\ \bibnamefont {Chakravadhanula}}, \bibinfo {author} {\bibfnamefont
  {C.}~\bibnamefont {K{\"u}bel}}, \bibinfo {author} {\bibfnamefont {S.~S.}\
  \bibnamefont {Bhattacharya}}, \bibinfo {author} {\bibfnamefont {A.~S.}\
  \bibnamefont {Gandhi}}, \ and\ \bibinfo {author} {\bibfnamefont
  {H.}~\bibnamefont {Hahn}},\ }\href@noop {} {\bibfield  {journal} {\bibinfo
  {journal} {Materials Research Letters}\ }\textbf {\bibinfo {volume} {5}},\
  \bibinfo {pages} {102} (\bibinfo {year} {2017})}\BibitemShut {NoStop}%
\bibitem [{\citenamefont {Jiang}\ \emph {et~al.}(2018)\citenamefont {Jiang},
  \citenamefont {Hu}, \citenamefont {Gild}, \citenamefont {Zhou}, \citenamefont
  {Nie}, \citenamefont {Qin}, \citenamefont {Harrington}, \citenamefont
  {Vecchio},\ and\ \citenamefont {Luo}}]{Jiang:2018p116}%
  \BibitemOpen
  \bibfield  {author} {\bibinfo {author} {\bibfnamefont {S.}~\bibnamefont
  {Jiang}}, \bibinfo {author} {\bibfnamefont {T.}~\bibnamefont {Hu}}, \bibinfo
  {author} {\bibfnamefont {J.}~\bibnamefont {Gild}}, \bibinfo {author}
  {\bibfnamefont {N.}~\bibnamefont {Zhou}}, \bibinfo {author} {\bibfnamefont
  {J.}~\bibnamefont {Nie}}, \bibinfo {author} {\bibfnamefont {M.}~\bibnamefont
  {Qin}}, \bibinfo {author} {\bibfnamefont {T.}~\bibnamefont {Harrington}},
  \bibinfo {author} {\bibfnamefont {K.}~\bibnamefont {Vecchio}}, \ and\
  \bibinfo {author} {\bibfnamefont {J.}~\bibnamefont {Luo}},\ }\href@noop {}
  {\bibfield  {journal} {\bibinfo  {journal} {Scripta Materialia}\ }\textbf
  {\bibinfo {volume} {142}},\ \bibinfo {pages} {116} (\bibinfo {year}
  {2018})}\BibitemShut {NoStop}%
\bibitem [{\citenamefont {Anand}\ \emph {et~al.}(2018)\citenamefont {Anand},
  \citenamefont {Wynn}, \citenamefont {Handley},\ and\ \citenamefont
  {Freeman}}]{Anand:2018p119}%
  \BibitemOpen
  \bibfield  {author} {\bibinfo {author} {\bibfnamefont {G.}~\bibnamefont
  {Anand}}, \bibinfo {author} {\bibfnamefont {A.~P.}\ \bibnamefont {Wynn}},
  \bibinfo {author} {\bibfnamefont {C.~M.}\ \bibnamefont {Handley}}, \ and\
  \bibinfo {author} {\bibfnamefont {C.~L.}\ \bibnamefont {Freeman}},\
  }\href@noop {} {\bibfield  {journal} {\bibinfo  {journal} {Acta Materialia}\
  }\textbf {\bibinfo {volume} {146}},\ \bibinfo {pages} {119} (\bibinfo {year}
  {2018})}\BibitemShut {NoStop}%
\bibitem [{\citenamefont {Sarkar}\ \emph {et~al.}(2018)\citenamefont {Sarkar},
  \citenamefont {Velasco}, \citenamefont {Wang}, \citenamefont {Wang},
  \citenamefont {Talasila}, \citenamefont {de~Biasi}, \citenamefont
  {K{\"u}bel}, \citenamefont {Brezesinski}, \citenamefont {Bhattacharya},
  \citenamefont {Hahn} \emph {et~al.}}]{Sarkar:2018p3400}%
  \BibitemOpen
  \bibfield  {author} {\bibinfo {author} {\bibfnamefont {A.}~\bibnamefont
  {Sarkar}}, \bibinfo {author} {\bibfnamefont {L.}~\bibnamefont {Velasco}},
  \bibinfo {author} {\bibfnamefont {D.}~\bibnamefont {Wang}}, \bibinfo {author}
  {\bibfnamefont {Q.}~\bibnamefont {Wang}}, \bibinfo {author} {\bibfnamefont
  {G.}~\bibnamefont {Talasila}}, \bibinfo {author} {\bibfnamefont
  {L.}~\bibnamefont {de~Biasi}}, \bibinfo {author} {\bibfnamefont
  {C.}~\bibnamefont {K{\"u}bel}}, \bibinfo {author} {\bibfnamefont
  {T.}~\bibnamefont {Brezesinski}}, \bibinfo {author} {\bibfnamefont {S.~S.}\
  \bibnamefont {Bhattacharya}}, \bibinfo {author} {\bibfnamefont
  {H.}~\bibnamefont {Hahn}},  \emph {et~al.},\ }\href@noop {} {\bibfield
  {journal} {\bibinfo  {journal} {Nature communications}\ }\textbf {\bibinfo
  {volume} {9}},\ \bibinfo {pages} {3400} (\bibinfo {year} {2018})}\BibitemShut
  {NoStop}%
\bibitem [{\citenamefont {Sharma}\ \emph {et~al.}(2018)\citenamefont {Sharma},
  \citenamefont {Musico}, \citenamefont {Gao}, \citenamefont {Hua},
  \citenamefont {May}, \citenamefont {Herklotz}, \citenamefont {Rastogi},
  \citenamefont {Mandrus}, \citenamefont {Yan}, \citenamefont {Lee} \emph
  {et~al.}}]{Sharma:2018p060404}%
  \BibitemOpen
  \bibfield  {author} {\bibinfo {author} {\bibfnamefont {Y.}~\bibnamefont
  {Sharma}}, \bibinfo {author} {\bibfnamefont {B.~L.}\ \bibnamefont {Musico}},
  \bibinfo {author} {\bibfnamefont {X.}~\bibnamefont {Gao}}, \bibinfo {author}
  {\bibfnamefont {C.}~\bibnamefont {Hua}}, \bibinfo {author} {\bibfnamefont
  {A.~F.}\ \bibnamefont {May}}, \bibinfo {author} {\bibfnamefont
  {A.}~\bibnamefont {Herklotz}}, \bibinfo {author} {\bibfnamefont
  {A.}~\bibnamefont {Rastogi}}, \bibinfo {author} {\bibfnamefont
  {D.}~\bibnamefont {Mandrus}}, \bibinfo {author} {\bibfnamefont
  {J.}~\bibnamefont {Yan}}, \bibinfo {author} {\bibfnamefont {H.~N.}\
  \bibnamefont {Lee}},  \emph {et~al.},\ }\href@noop {} {\bibfield  {journal}
  {\bibinfo  {journal} {Physical Review Materials}\ }\textbf {\bibinfo {volume}
  {2}},\ \bibinfo {pages} {060404} (\bibinfo {year} {2018})}\BibitemShut
  {NoStop}%
\bibitem [{\citenamefont {Dkabrowa}\ \emph {et~al.}(2018)\citenamefont
  {Dkabrowa}, \citenamefont {Stygar}, \citenamefont {Mikua}, \citenamefont
  {Knapik}, \citenamefont {Mroczka}, \citenamefont {Tejchman}, \citenamefont
  {Danielewski},\ and\ \citenamefont {Martin}}]{Dkabrowa:2018p32}%
  \BibitemOpen
  \bibfield  {author} {\bibinfo {author} {\bibfnamefont {J.}~\bibnamefont
  {Dkabrowa}}, \bibinfo {author} {\bibfnamefont {M.}~\bibnamefont {Stygar}},
  \bibinfo {author} {\bibfnamefont {A.}~\bibnamefont {Mikua}}, \bibinfo
  {author} {\bibfnamefont {A.}~\bibnamefont {Knapik}}, \bibinfo {author}
  {\bibfnamefont {K.}~\bibnamefont {Mroczka}}, \bibinfo {author} {\bibfnamefont
  {W.}~\bibnamefont {Tejchman}}, \bibinfo {author} {\bibfnamefont
  {M.}~\bibnamefont {Danielewski}}, \ and\ \bibinfo {author} {\bibfnamefont
  {M.}~\bibnamefont {Martin}},\ }\href@noop {} {\bibfield  {journal} {\bibinfo
  {journal} {Materials Letters}\ }\textbf {\bibinfo {volume} {216}},\ \bibinfo
  {pages} {32} (\bibinfo {year} {2018})}\BibitemShut {NoStop}%
\bibitem [{\citenamefont {Witte}\ \emph {et~al.}(2019)\citenamefont {Witte},
  \citenamefont {Sarkar}, \citenamefont {Kruk}, \citenamefont {Eggert},
  \citenamefont {Brand}, \citenamefont {Wende},\ and\ \citenamefont
  {Hahn}}]{Witte:2019p034406}%
  \BibitemOpen
  \bibfield  {author} {\bibinfo {author} {\bibfnamefont {R.}~\bibnamefont
  {Witte}}, \bibinfo {author} {\bibfnamefont {A.}~\bibnamefont {Sarkar}},
  \bibinfo {author} {\bibfnamefont {R.}~\bibnamefont {Kruk}}, \bibinfo {author}
  {\bibfnamefont {B.}~\bibnamefont {Eggert}}, \bibinfo {author} {\bibfnamefont
  {R.~A.}\ \bibnamefont {Brand}}, \bibinfo {author} {\bibfnamefont
  {H.}~\bibnamefont {Wende}}, \ and\ \bibinfo {author} {\bibfnamefont
  {H.}~\bibnamefont {Hahn}},\ }\href@noop {} {\bibfield  {journal} {\bibinfo
  {journal} {Physical Review Materials}\ }\textbf {\bibinfo {volume} {3}},\
  \bibinfo {pages} {034406} (\bibinfo {year} {2019})}\BibitemShut {NoStop}%
\bibitem [{\citenamefont {Sarkar}\ \emph {et~al.}(2019)\citenamefont {Sarkar},
  \citenamefont {Wang}, \citenamefont {Schiele}, \citenamefont {Chellali},
  \citenamefont {Bhattacharya}, \citenamefont {Wang}, \citenamefont
  {Brezesinski}, \citenamefont {Hahn}, \citenamefont {Velasco},\ and\
  \citenamefont {Breitung}}]{Sarkar:2019p1806236}%
  \BibitemOpen
  \bibfield  {author} {\bibinfo {author} {\bibfnamefont {A.}~\bibnamefont
  {Sarkar}}, \bibinfo {author} {\bibfnamefont {Q.}~\bibnamefont {Wang}},
  \bibinfo {author} {\bibfnamefont {A.}~\bibnamefont {Schiele}}, \bibinfo
  {author} {\bibfnamefont {M.~R.}\ \bibnamefont {Chellali}}, \bibinfo {author}
  {\bibfnamefont {S.~S.}\ \bibnamefont {Bhattacharya}}, \bibinfo {author}
  {\bibfnamefont {D.}~\bibnamefont {Wang}}, \bibinfo {author} {\bibfnamefont
  {T.}~\bibnamefont {Brezesinski}}, \bibinfo {author} {\bibfnamefont
  {H.}~\bibnamefont {Hahn}}, \bibinfo {author} {\bibfnamefont {L.}~\bibnamefont
  {Velasco}}, \ and\ \bibinfo {author} {\bibfnamefont {B.}~\bibnamefont
  {Breitung}},\ }\href@noop {} {\bibfield  {journal} {\bibinfo  {journal}
  {Advanced Materials}\ }\textbf {\bibinfo {volume} {31}},\ \bibinfo {pages}
  {1806236} (\bibinfo {year} {2019})}\BibitemShut {NoStop}%
\bibitem [{\citenamefont {Zhang}\ \emph {et~al.}(2019)\citenamefont {Zhang},
  \citenamefont {Yan}, \citenamefont {Calder}, \citenamefont {Zheng},
  \citenamefont {McGuire}, \citenamefont {Abernathy}, \citenamefont {Ren},
  \citenamefont {Lapidus}, \citenamefont {Page}, \citenamefont {Zheng} \emph
  {et~al.}}]{Zhang2019plong}%
  \BibitemOpen
  \bibfield  {author} {\bibinfo {author} {\bibfnamefont {J.}~\bibnamefont
  {Zhang}}, \bibinfo {author} {\bibfnamefont {J.}~\bibnamefont {Yan}}, \bibinfo
  {author} {\bibfnamefont {S.}~\bibnamefont {Calder}}, \bibinfo {author}
  {\bibfnamefont {Q.}~\bibnamefont {Zheng}}, \bibinfo {author} {\bibfnamefont
  {M.~A.}\ \bibnamefont {McGuire}}, \bibinfo {author} {\bibfnamefont {D.~L.}\
  \bibnamefont {Abernathy}}, \bibinfo {author} {\bibfnamefont {Y.}~\bibnamefont
  {Ren}}, \bibinfo {author} {\bibfnamefont {S.~H.}\ \bibnamefont {Lapidus}},
  \bibinfo {author} {\bibfnamefont {K.}~\bibnamefont {Page}}, \bibinfo {author}
  {\bibfnamefont {H.}~\bibnamefont {Zheng}},  \emph {et~al.},\ }\href@noop {}
  {\bibfield  {journal} {\bibinfo  {journal} {Chemistry of Materials}\ }
  (\bibinfo {year} {2019})}\BibitemShut {NoStop}%
\bibitem [{\citenamefont {Meisenheimer}, \citenamefont {Kratofil},\ and\
  \citenamefont {Heron}(2017)}]{Meisenheimer:2017p13344}%
  \BibitemOpen
  \bibfield  {author} {\bibinfo {author} {\bibfnamefont {P.}~\bibnamefont
  {Meisenheimer}}, \bibinfo {author} {\bibfnamefont {T.}~\bibnamefont
  {Kratofil}}, \ and\ \bibinfo {author} {\bibfnamefont {J.}~\bibnamefont
  {Heron}},\ }\href@noop {} {\bibfield  {journal} {\bibinfo  {journal}
  {Scientific reports}\ }\textbf {\bibinfo {volume} {7}},\ \bibinfo {pages}
  {13344} (\bibinfo {year} {2017})}\BibitemShut {NoStop}%
\bibitem [{\citenamefont {Sharma}\ \emph {et~al.}(2019)\citenamefont {Sharma},
  \citenamefont {Zheng}, \citenamefont {Mazza}, \citenamefont {Skoropata},
  \citenamefont {Heitmann}, \citenamefont {Gai}, \citenamefont {Musico},
  \citenamefont {Miceli}, \citenamefont {Sales}, \citenamefont {Keppens} \emph
  {et~al.}}]{sharma:2019magnetic}%
  \BibitemOpen
  \bibfield  {author} {\bibinfo {author} {\bibfnamefont {Y.}~\bibnamefont
  {Sharma}}, \bibinfo {author} {\bibfnamefont {Q.}~\bibnamefont {Zheng}},
  \bibinfo {author} {\bibfnamefont {A.~R.}\ \bibnamefont {Mazza}}, \bibinfo
  {author} {\bibfnamefont {E.}~\bibnamefont {Skoropata}}, \bibinfo {author}
  {\bibfnamefont {T.}~\bibnamefont {Heitmann}}, \bibinfo {author}
  {\bibfnamefont {Z.}~\bibnamefont {Gai}}, \bibinfo {author} {\bibfnamefont
  {B.}~\bibnamefont {Musico}}, \bibinfo {author} {\bibfnamefont {P.~F.}\
  \bibnamefont {Miceli}}, \bibinfo {author} {\bibfnamefont {B.~C.}\
  \bibnamefont {Sales}}, \bibinfo {author} {\bibfnamefont {V.}~\bibnamefont
  {Keppens}},  \emph {et~al.},\ }\href@noop {} {\bibfield  {journal} {\bibinfo
  {journal} {arXiv preprint arXiv:1909.05019}\ } (\bibinfo {year}
  {2019})}\BibitemShut {NoStop}%
\bibitem [{\citenamefont {Medarde}(1997)}]{Medarde:1997p1679}%
  \BibitemOpen
  \bibfield  {author} {\bibinfo {author} {\bibfnamefont {M.~L.}\ \bibnamefont
  {Medarde}},\ }\href@noop {} {\bibfield  {journal} {\bibinfo  {journal}
  {Journal of Physics: Condensed Matter}\ }\textbf {\bibinfo {volume} {9}},\
  \bibinfo {pages} {1679} (\bibinfo {year} {1997})}\BibitemShut {NoStop}%
\bibitem [{\citenamefont {Catalan}(2008)}]{Catalan:2008p729}%
  \BibitemOpen
  \bibfield  {author} {\bibinfo {author} {\bibfnamefont {G.}~\bibnamefont
  {Catalan}},\ }\href {\doibase 10.1080/01411590801992463} {\bibfield
  {journal} {\bibinfo  {journal} {Phase Transitions}\ }\textbf {\bibinfo
  {volume} {81}},\ \bibinfo {pages} {729} (\bibinfo {year} {2008})}\BibitemShut
  {NoStop}%
\bibitem [{\citenamefont {Staub}\ \emph {et~al.}(2002)\citenamefont {Staub},
  \citenamefont {Meijer}, \citenamefont {Fauth}, \citenamefont {Allenspach},
  \citenamefont {Bednorz}, \citenamefont {Karpinski}, \citenamefont {Kazakov},
  \citenamefont {Paolasini},\ and\ \citenamefont
  {d'Acapito}}]{Staub:2002p126402}%
  \BibitemOpen
  \bibfield  {author} {\bibinfo {author} {\bibfnamefont {U.}~\bibnamefont
  {Staub}}, \bibinfo {author} {\bibfnamefont {G.~I.}\ \bibnamefont {Meijer}},
  \bibinfo {author} {\bibfnamefont {F.}~\bibnamefont {Fauth}}, \bibinfo
  {author} {\bibfnamefont {R.}~\bibnamefont {Allenspach}}, \bibinfo {author}
  {\bibfnamefont {J.~G.}\ \bibnamefont {Bednorz}}, \bibinfo {author}
  {\bibfnamefont {J.}~\bibnamefont {Karpinski}}, \bibinfo {author}
  {\bibfnamefont {S.~M.}\ \bibnamefont {Kazakov}}, \bibinfo {author}
  {\bibfnamefont {L.}~\bibnamefont {Paolasini}}, \ and\ \bibinfo {author}
  {\bibfnamefont {F.}~\bibnamefont {d'Acapito}},\ }\href {\doibase
  10.1103/PhysRevLett.88.126402} {\bibfield  {journal} {\bibinfo  {journal}
  {Phys. Rev. Lett.}\ }\textbf {\bibinfo {volume} {88}},\ \bibinfo {pages}
  {126402} (\bibinfo {year} {2002})}\BibitemShut {NoStop}%
\bibitem [{\citenamefont {Middey}\ \emph {et~al.}(2016)\citenamefont {Middey},
  \citenamefont {Chakhalian}, \citenamefont {Mahadevan}, \citenamefont
  {Freeland}, \citenamefont {Millis},\ and\ \citenamefont
  {Sarma}}]{Middey:2016p305}%
  \BibitemOpen
  \bibfield  {author} {\bibinfo {author} {\bibfnamefont {S.}~\bibnamefont
  {Middey}}, \bibinfo {author} {\bibfnamefont {J.}~\bibnamefont {Chakhalian}},
  \bibinfo {author} {\bibfnamefont {P.}~\bibnamefont {Mahadevan}}, \bibinfo
  {author} {\bibfnamefont {J.~W.}\ \bibnamefont {Freeland}}, \bibinfo {author}
  {\bibfnamefont {A.~J.}\ \bibnamefont {Millis}}, \ and\ \bibinfo {author}
  {\bibfnamefont {D.~D.}\ \bibnamefont {Sarma}},\ }\href {\doibase
  10.1146/annurev-matsci-070115-032057} {\bibfield  {journal} {\bibinfo
  {journal} {Annual Review of Materials Research}\ }\textbf {\bibinfo {volume}
  {46}},\ \bibinfo {pages} {305} (\bibinfo {year} {2016})}\BibitemShut
  {NoStop}%
\bibitem [{\citenamefont {Catalano}\ \emph {et~al.}(2018)\citenamefont
  {Catalano}, \citenamefont {Gibert}, \citenamefont {Fowlie}, \citenamefont
  {{\'I}{\~n}iguez}, \citenamefont {Triscone},\ and\ \citenamefont
  {Kreisel}}]{Catalano:2018p046501}%
  \BibitemOpen
  \bibfield  {author} {\bibinfo {author} {\bibfnamefont {S.}~\bibnamefont
  {Catalano}}, \bibinfo {author} {\bibfnamefont {M.}~\bibnamefont {Gibert}},
  \bibinfo {author} {\bibfnamefont {J.}~\bibnamefont {Fowlie}}, \bibinfo
  {author} {\bibfnamefont {J.}~\bibnamefont {{\'I}{\~n}iguez}}, \bibinfo
  {author} {\bibfnamefont {J.-M.}\ \bibnamefont {Triscone}}, \ and\ \bibinfo
  {author} {\bibfnamefont {J.}~\bibnamefont {Kreisel}},\ }\href {\doibase
  10.1088/1361-6633/aaa37a} {\bibfield  {journal} {\bibinfo  {journal} {Reports
  on Progress in Physics}\ }\textbf {\bibinfo {volume} {81}},\ \bibinfo {pages}
  {046501} (\bibinfo {year} {2018})}\BibitemShut {NoStop}%
\bibitem [{\citenamefont {Ha}\ \emph {et~al.}(2012)\citenamefont {Ha},
  \citenamefont {Otaki}, \citenamefont {Jaramillo}, \citenamefont {Podpirka},\
  and\ \citenamefont {Ramanathan}}]{Ha:2012p233}%
  \BibitemOpen
  \bibfield  {author} {\bibinfo {author} {\bibfnamefont {S.~D.}\ \bibnamefont
  {Ha}}, \bibinfo {author} {\bibfnamefont {M.}~\bibnamefont {Otaki}}, \bibinfo
  {author} {\bibfnamefont {R.}~\bibnamefont {Jaramillo}}, \bibinfo {author}
  {\bibfnamefont {A.}~\bibnamefont {Podpirka}}, \ and\ \bibinfo {author}
  {\bibfnamefont {S.}~\bibnamefont {Ramanathan}},\ }\href {\doibase
  http://dx.doi.org/10.1016/j.jssc.2012.02.047} {\bibfield  {journal} {\bibinfo
   {journal} {Journal of Solid State Chemistry}\ }\textbf {\bibinfo {volume}
  {190}},\ \bibinfo {pages} {233 } (\bibinfo {year} {2012})}\BibitemShut
  {NoStop}%
\bibitem [{\citenamefont {Liu}\ \emph {et~al.}(2013)\citenamefont {Liu},
  \citenamefont {Kargarian}, \citenamefont {Kareev}, \citenamefont {Gray},
  \citenamefont {Ryan}, \citenamefont {Cruz}, \citenamefont {Tahir},
  \citenamefont {Chuang}, \citenamefont {Guo}, \citenamefont {Rondinelli},
  \citenamefont {Freeland}, \citenamefont {Fiete},\ and\ \citenamefont
  {Chakhalian}}]{Liu:2013p2714}%
  \BibitemOpen
  \bibfield  {author} {\bibinfo {author} {\bibfnamefont {J.}~\bibnamefont
  {Liu}}, \bibinfo {author} {\bibfnamefont {M.}~\bibnamefont {Kargarian}},
  \bibinfo {author} {\bibfnamefont {M.}~\bibnamefont {Kareev}}, \bibinfo
  {author} {\bibfnamefont {B.}~\bibnamefont {Gray}}, \bibinfo {author}
  {\bibfnamefont {P.~J.}\ \bibnamefont {Ryan}}, \bibinfo {author}
  {\bibfnamefont {A.}~\bibnamefont {Cruz}}, \bibinfo {author} {\bibfnamefont
  {N.}~\bibnamefont {Tahir}}, \bibinfo {author} {\bibfnamefont {Y.-D.}\
  \bibnamefont {Chuang}}, \bibinfo {author} {\bibfnamefont {J.}~\bibnamefont
  {Guo}}, \bibinfo {author} {\bibfnamefont {J.~M.}\ \bibnamefont {Rondinelli}},
  \bibinfo {author} {\bibfnamefont {J.~W.}\ \bibnamefont {Freeland}}, \bibinfo
  {author} {\bibfnamefont {G.~A.}\ \bibnamefont {Fiete}}, \ and\ \bibinfo
  {author} {\bibfnamefont {J.}~\bibnamefont {Chakhalian}},\ }\href
  {http://dx.doi.org/10.1038/ncomms3714} {\bibfield  {journal} {\bibinfo
  {journal} {Nat Commun}\ }\textbf {\bibinfo {volume} {4}},\ \bibinfo {pages}
  {2714} (\bibinfo {year} {2013})}\BibitemShut {NoStop}%
\bibitem [{\citenamefont {Feigl}\ \emph {et~al.}(2013)\citenamefont {Feigl},
  \citenamefont {Schultz}, \citenamefont {Ohya}, \citenamefont {Ouellette},
  \citenamefont {Kozhanov},\ and\ \citenamefont {Palmstram}}]{Feigl:2013p51}%
  \BibitemOpen
  \bibfield  {author} {\bibinfo {author} {\bibfnamefont {L.}~\bibnamefont
  {Feigl}}, \bibinfo {author} {\bibfnamefont {B.}~\bibnamefont {Schultz}},
  \bibinfo {author} {\bibfnamefont {S.}~\bibnamefont {Ohya}}, \bibinfo {author}
  {\bibfnamefont {D.}~\bibnamefont {Ouellette}}, \bibinfo {author}
  {\bibfnamefont {A.}~\bibnamefont {Kozhanov}}, \ and\ \bibinfo {author}
  {\bibfnamefont {C.}~\bibnamefont {Palmstram}},\ }\href {\doibase
  http://dx.doi.org/10.1016/j.jcrysgro.2012.12.018} {\bibfield  {journal}
  {\bibinfo  {journal} {Journal of Crystal Growth}\ }\textbf {\bibinfo {volume}
  {366}},\ \bibinfo {pages} {51 } (\bibinfo {year} {2013})}\BibitemShut
  {NoStop}%
\bibitem [{\citenamefont {Meyers}\ \emph {et~al.}(2013)\citenamefont {Meyers},
  \citenamefont {Middey}, \citenamefont {Kareev}, \citenamefont {van
  Veenendaal}, \citenamefont {Moon}, \citenamefont {Gray}, \citenamefont {Liu},
  \citenamefont {Freeland},\ and\ \citenamefont
  {Chakhalian}}]{Meyers:2013p075116}%
  \BibitemOpen
  \bibfield  {author} {\bibinfo {author} {\bibfnamefont {D.}~\bibnamefont
  {Meyers}}, \bibinfo {author} {\bibfnamefont {S.}~\bibnamefont {Middey}},
  \bibinfo {author} {\bibfnamefont {M.}~\bibnamefont {Kareev}}, \bibinfo
  {author} {\bibfnamefont {M.}~\bibnamefont {van Veenendaal}}, \bibinfo
  {author} {\bibfnamefont {E.~J.}\ \bibnamefont {Moon}}, \bibinfo {author}
  {\bibfnamefont {B.~A.}\ \bibnamefont {Gray}}, \bibinfo {author}
  {\bibfnamefont {J.}~\bibnamefont {Liu}}, \bibinfo {author} {\bibfnamefont
  {J.~W.}\ \bibnamefont {Freeland}}, \ and\ \bibinfo {author} {\bibfnamefont
  {J.}~\bibnamefont {Chakhalian}},\ }\href {\doibase
  10.1103/PhysRevB.88.075116} {\bibfield  {journal} {\bibinfo  {journal} {Phys.
  Rev. B}\ }\textbf {\bibinfo {volume} {88}},\ \bibinfo {pages} {075116}
  (\bibinfo {year} {2013})}\BibitemShut {NoStop}%
\bibitem [{\citenamefont {Mikheev}\ \emph {et~al.}(2015)\citenamefont
  {Mikheev}, \citenamefont {Hauser}, \citenamefont {Himmetoglu}, \citenamefont
  {Moreno}, \citenamefont {Janotti}, \citenamefont {Van~de Walle},\ and\
  \citenamefont {Stemmer}}]{Mikheev:2015p1500797}%
  \BibitemOpen
  \bibfield  {author} {\bibinfo {author} {\bibfnamefont {E.}~\bibnamefont
  {Mikheev}}, \bibinfo {author} {\bibfnamefont {A.~J.}\ \bibnamefont {Hauser}},
  \bibinfo {author} {\bibfnamefont {B.}~\bibnamefont {Himmetoglu}}, \bibinfo
  {author} {\bibfnamefont {N.~E.}\ \bibnamefont {Moreno}}, \bibinfo {author}
  {\bibfnamefont {A.}~\bibnamefont {Janotti}}, \bibinfo {author} {\bibfnamefont
  {C.~G.}\ \bibnamefont {Van~de Walle}}, \ and\ \bibinfo {author}
  {\bibfnamefont {S.}~\bibnamefont {Stemmer}},\ }\href {\doibase
  10.1126/sciadv.1500797} {\bibfield  {journal} {\bibinfo  {journal} {Science
  Advances}\ }\textbf {\bibinfo {volume} {1}},\ \bibinfo {pages} {e1500797}
  (\bibinfo {year} {2015})},\ \Eprint
  {http://arxiv.org/abs/http://advances.sciencemag.org/content/1/10/e1500797.full.pdf}
  {http://advances.sciencemag.org/content/1/10/e1500797.full.pdf} \BibitemShut
  {NoStop}%
\bibitem [{\citenamefont {Hepting}\ \emph {et~al.}(2014)\citenamefont
  {Hepting}, \citenamefont {Minola}, \citenamefont {Frano}, \citenamefont
  {Cristiani}, \citenamefont {Logvenov}, \citenamefont {Schierle},
  \citenamefont {Wu}, \citenamefont {Bluschke}, \citenamefont {Weschke},
  \citenamefont {Habermeier}, \citenamefont {Benckiser}, \citenamefont
  {Le~Tacon},\ and\ \citenamefont {Keimer}}]{Hepting:2014p227206}%
  \BibitemOpen
  \bibfield  {author} {\bibinfo {author} {\bibfnamefont {M.}~\bibnamefont
  {Hepting}}, \bibinfo {author} {\bibfnamefont {M.}~\bibnamefont {Minola}},
  \bibinfo {author} {\bibfnamefont {A.}~\bibnamefont {Frano}}, \bibinfo
  {author} {\bibfnamefont {G.}~\bibnamefont {Cristiani}}, \bibinfo {author}
  {\bibfnamefont {G.}~\bibnamefont {Logvenov}}, \bibinfo {author}
  {\bibfnamefont {E.}~\bibnamefont {Schierle}}, \bibinfo {author}
  {\bibfnamefont {M.}~\bibnamefont {Wu}}, \bibinfo {author} {\bibfnamefont
  {M.}~\bibnamefont {Bluschke}}, \bibinfo {author} {\bibfnamefont
  {E.}~\bibnamefont {Weschke}}, \bibinfo {author} {\bibfnamefont {H.-U.}\
  \bibnamefont {Habermeier}}, \bibinfo {author} {\bibfnamefont
  {E.}~\bibnamefont {Benckiser}}, \bibinfo {author} {\bibfnamefont
  {M.}~\bibnamefont {Le~Tacon}}, \ and\ \bibinfo {author} {\bibfnamefont
  {B.}~\bibnamefont {Keimer}},\ }\href {\doibase
  10.1103/PhysRevLett.113.227206} {\bibfield  {journal} {\bibinfo  {journal}
  {Phys. Rev. Lett.}\ }\textbf {\bibinfo {volume} {113}},\ \bibinfo {pages}
  {227206} (\bibinfo {year} {2014})}\BibitemShut {NoStop}%
\bibitem [{\citenamefont {Scherwitzl}\ \emph {et~al.}(2011)\citenamefont
  {Scherwitzl}, \citenamefont {Gariglio}, \citenamefont {Gabay}, \citenamefont
  {Zubko}, \citenamefont {Gibert},\ and\ \citenamefont
  {Triscone}}]{Scherwitzl:2011p246403}%
  \BibitemOpen
  \bibfield  {author} {\bibinfo {author} {\bibfnamefont {R.}~\bibnamefont
  {Scherwitzl}}, \bibinfo {author} {\bibfnamefont {S.}~\bibnamefont
  {Gariglio}}, \bibinfo {author} {\bibfnamefont {M.}~\bibnamefont {Gabay}},
  \bibinfo {author} {\bibfnamefont {P.}~\bibnamefont {Zubko}}, \bibinfo
  {author} {\bibfnamefont {M.}~\bibnamefont {Gibert}}, \ and\ \bibinfo {author}
  {\bibfnamefont {J.-M.}\ \bibnamefont {Triscone}},\ }\href {\doibase
  10.1103/PhysRevLett.106.246403} {\bibfield  {journal} {\bibinfo  {journal}
  {Phys. Rev. Lett.}\ }\textbf {\bibinfo {volume} {106}},\ \bibinfo {pages}
  {246403} (\bibinfo {year} {2011})}\BibitemShut {NoStop}%
\bibitem [{\citenamefont {Bruno}\ \emph {et~al.}(2013)\citenamefont {Bruno},
  \citenamefont {Rushchanskii}, \citenamefont {Valencia}, \citenamefont
  {Dumont}, \citenamefont {Carr\'et\'ero}, \citenamefont {Jacquet},
  \citenamefont {Abrudan}, \citenamefont {Bl\"ugel}, \citenamefont {Le\ifmmode
  \check{z}\else \v{z}\fi{}ai\ifmmode~\acute{c}\else \'{c}\fi{}}, \citenamefont
  {Bibes},\ and\ \citenamefont {Barth\'el\'emy}}]{Bruno:2013p195108}%
  \BibitemOpen
  \bibfield  {author} {\bibinfo {author} {\bibfnamefont {F.~Y.}\ \bibnamefont
  {Bruno}}, \bibinfo {author} {\bibfnamefont {K.~Z.}\ \bibnamefont
  {Rushchanskii}}, \bibinfo {author} {\bibfnamefont {S.}~\bibnamefont
  {Valencia}}, \bibinfo {author} {\bibfnamefont {Y.}~\bibnamefont {Dumont}},
  \bibinfo {author} {\bibfnamefont {C.}~\bibnamefont {Carr\'et\'ero}}, \bibinfo
  {author} {\bibfnamefont {E.}~\bibnamefont {Jacquet}}, \bibinfo {author}
  {\bibfnamefont {R.}~\bibnamefont {Abrudan}}, \bibinfo {author} {\bibfnamefont
  {S.}~\bibnamefont {Bl\"ugel}}, \bibinfo {author} {\bibfnamefont
  {M.}~\bibnamefont {Le\ifmmode \check{z}\else
  \v{z}\fi{}ai\ifmmode~\acute{c}\else \'{c}\fi{}}}, \bibinfo {author}
  {\bibfnamefont {M.}~\bibnamefont {Bibes}}, \ and\ \bibinfo {author}
  {\bibfnamefont {A.}~\bibnamefont {Barth\'el\'emy}},\ }\href {\doibase
  10.1103/PhysRevB.88.195108} {\bibfield  {journal} {\bibinfo  {journal} {Phys.
  Rev. B}\ }\textbf {\bibinfo {volume} {88}},\ \bibinfo {pages} {195108}
  (\bibinfo {year} {2013})}\BibitemShut {NoStop}%
\bibitem [{\citenamefont {Garc\'{\i}a-Mu\~noz}\ \emph
  {et~al.}(1992)\citenamefont {Garc\'{\i}a-Mu\~noz}, \citenamefont
  {Rodr\'{\i}guez-Carvajal}, \citenamefont {Lacorre},\ and\ \citenamefont
  {Torrance}}]{Garica:1992p4414}%
  \BibitemOpen
  \bibfield  {author} {\bibinfo {author} {\bibfnamefont {J.~L.}\ \bibnamefont
  {Garc\'{\i}a-Mu\~noz}}, \bibinfo {author} {\bibfnamefont {J.}~\bibnamefont
  {Rodr\'{\i}guez-Carvajal}}, \bibinfo {author} {\bibfnamefont
  {P.}~\bibnamefont {Lacorre}}, \ and\ \bibinfo {author} {\bibfnamefont
  {J.~B.}\ \bibnamefont {Torrance}},\ }\href {\doibase
  10.1103/PhysRevB.46.4414} {\bibfield  {journal} {\bibinfo  {journal} {Phys.
  Rev. B}\ }\textbf {\bibinfo {volume} {46}},\ \bibinfo {pages} {4414}
  (\bibinfo {year} {1992})}\BibitemShut {NoStop}%
\bibitem [{\citenamefont {Alonso}\ \emph {et~al.}(1999)\citenamefont {Alonso},
  \citenamefont {Mart{\'\i}nez-Lope}, \citenamefont {Casais}, \citenamefont
  {Aranda},\ and\ \citenamefont {Fern{\'a}ndez-D{\'\i}az}}]{Alonso:1999p4754}%
  \BibitemOpen
  \bibfield  {author} {\bibinfo {author} {\bibfnamefont {J.~A.}\ \bibnamefont
  {Alonso}}, \bibinfo {author} {\bibfnamefont {M.~J.}\ \bibnamefont
  {Mart{\'\i}nez-Lope}}, \bibinfo {author} {\bibfnamefont {M.~T.}\ \bibnamefont
  {Casais}}, \bibinfo {author} {\bibfnamefont {M.~A.~G.}\ \bibnamefont
  {Aranda}}, \ and\ \bibinfo {author} {\bibfnamefont {M.~T.}\ \bibnamefont
  {Fern{\'a}ndez-D{\'\i}az}},\ }\href {\doibase 10.1021/ja984015x} {\bibfield
  {journal} {\bibinfo  {journal} {Journal of the American Chemical Society}\
  }\textbf {\bibinfo {volume} {121}},\ \bibinfo {pages} {4754} (\bibinfo {year}
  {1999})},\ \Eprint {http://arxiv.org/abs/http://dx.doi.org/10.1021/ja984015x}
  {http://dx.doi.org/10.1021/ja984015x} \BibitemShut {NoStop}%
\bibitem [{\citenamefont {Zhou}, \citenamefont {Goodenough},\ and\
  \citenamefont {Dabrowski}(2003)}]{Zhou:2003p020404}%
  \BibitemOpen
  \bibfield  {author} {\bibinfo {author} {\bibfnamefont {J.-S.}\ \bibnamefont
  {Zhou}}, \bibinfo {author} {\bibfnamefont {J.}~\bibnamefont {Goodenough}}, \
  and\ \bibinfo {author} {\bibfnamefont {B.}~\bibnamefont {Dabrowski}},\
  }\href@noop {} {\bibfield  {journal} {\bibinfo  {journal} {Physical Review
  B}\ }\textbf {\bibinfo {volume} {67}},\ \bibinfo {pages} {020404} (\bibinfo
  {year} {2003})}\BibitemShut {NoStop}%
\bibitem [{\citenamefont {Ojha}\ \emph {et~al.}(2019)\citenamefont {Ojha},
  \citenamefont {Ray}, \citenamefont {Das}, \citenamefont {Middey},
  \citenamefont {Sarkar}, \citenamefont {Mahadevan}, \citenamefont {Wang},
  \citenamefont {Zhu}, \citenamefont {Liu}, \citenamefont {Kareev},\ and\
  \citenamefont {Chakhalian}}]{Ojha:2019p235153}%
  \BibitemOpen
  \bibfield  {author} {\bibinfo {author} {\bibfnamefont {S.~K.}\ \bibnamefont
  {Ojha}}, \bibinfo {author} {\bibfnamefont {S.}~\bibnamefont {Ray}}, \bibinfo
  {author} {\bibfnamefont {T.}~\bibnamefont {Das}}, \bibinfo {author}
  {\bibfnamefont {S.}~\bibnamefont {Middey}}, \bibinfo {author} {\bibfnamefont
  {S.}~\bibnamefont {Sarkar}}, \bibinfo {author} {\bibfnamefont
  {P.}~\bibnamefont {Mahadevan}}, \bibinfo {author} {\bibfnamefont
  {Z.}~\bibnamefont {Wang}}, \bibinfo {author} {\bibfnamefont {Y.}~\bibnamefont
  {Zhu}}, \bibinfo {author} {\bibfnamefont {X.}~\bibnamefont {Liu}}, \bibinfo
  {author} {\bibfnamefont {M.}~\bibnamefont {Kareev}}, \ and\ \bibinfo {author}
  {\bibfnamefont {J.}~\bibnamefont {Chakhalian}},\ }\href {\doibase
  10.1103/PhysRevB.99.235153} {\bibfield  {journal} {\bibinfo  {journal} {Phys.
  Rev. B}\ }\textbf {\bibinfo {volume} {99}},\ \bibinfo {pages} {235153}
  (\bibinfo {year} {2019})}\BibitemShut {NoStop}%
\bibitem [{\citenamefont {Garc\'{i}a-Mu\~noz}, \citenamefont
  {Rodr\'{i}guez-Carvajal},\ and\ \citenamefont
  {Lacorre}(1994)}]{Garcia:1994p978}%
  \BibitemOpen
  \bibfield  {author} {\bibinfo {author} {\bibfnamefont {J.~L.}\ \bibnamefont
  {Garc\'{i}a-Mu\~noz}}, \bibinfo {author} {\bibfnamefont {J.}~\bibnamefont
  {Rodr\'{i}guez-Carvajal}}, \ and\ \bibinfo {author} {\bibfnamefont
  {P.}~\bibnamefont {Lacorre}},\ }\href {\doibase 10.1103/PhysRevB.50.978}
  {\bibfield  {journal} {\bibinfo  {journal} {Phys. Rev. B}\ }\textbf {\bibinfo
  {volume} {50}},\ \bibinfo {pages} {978} (\bibinfo {year} {1994})}\BibitemShut
  {NoStop}%
\bibitem [{\citenamefont {Scagnoli}\ \emph {et~al.}(2006)\citenamefont
  {Scagnoli}, \citenamefont {Staub}, \citenamefont {Mulders}, \citenamefont
  {Janousch}, \citenamefont {Meijer}, \citenamefont {Hammerl}, \citenamefont
  {Tonnerre},\ and\ \citenamefont {Stojic}}]{Scagnoli:2006p100409}%
  \BibitemOpen
  \bibfield  {author} {\bibinfo {author} {\bibfnamefont {V.}~\bibnamefont
  {Scagnoli}}, \bibinfo {author} {\bibfnamefont {U.}~\bibnamefont {Staub}},
  \bibinfo {author} {\bibfnamefont {A.~M.}\ \bibnamefont {Mulders}}, \bibinfo
  {author} {\bibfnamefont {M.}~\bibnamefont {Janousch}}, \bibinfo {author}
  {\bibfnamefont {G.~I.}\ \bibnamefont {Meijer}}, \bibinfo {author}
  {\bibfnamefont {G.}~\bibnamefont {Hammerl}}, \bibinfo {author} {\bibfnamefont
  {J.~M.}\ \bibnamefont {Tonnerre}}, \ and\ \bibinfo {author} {\bibfnamefont
  {N.}~\bibnamefont {Stojic}},\ }\href {\doibase 10.1103/PhysRevB.73.100409}
  {\bibfield  {journal} {\bibinfo  {journal} {Phys. Rev. B}\ }\textbf {\bibinfo
  {volume} {73}},\ \bibinfo {pages} {100409} (\bibinfo {year}
  {2006})}\BibitemShut {NoStop}%
\bibitem [{\citenamefont {Middey}\ \emph
  {et~al.}(2018{\natexlab{a}})\citenamefont {Middey}, \citenamefont {Meyers},
  \citenamefont {Kumar~Patel}, \citenamefont {Liu}, \citenamefont {Kareev},
  \citenamefont {Shafer}, \citenamefont {Kim}, \citenamefont {Ryan},\ and\
  \citenamefont {Chakhalian}}]{Middey:2018p081602}%
  \BibitemOpen
  \bibfield  {author} {\bibinfo {author} {\bibfnamefont {S.}~\bibnamefont
  {Middey}}, \bibinfo {author} {\bibfnamefont {D.}~\bibnamefont {Meyers}},
  \bibinfo {author} {\bibfnamefont {R.}~\bibnamefont {Kumar~Patel}}, \bibinfo
  {author} {\bibfnamefont {X.}~\bibnamefont {Liu}}, \bibinfo {author}
  {\bibfnamefont {M.}~\bibnamefont {Kareev}}, \bibinfo {author} {\bibfnamefont
  {P.}~\bibnamefont {Shafer}}, \bibinfo {author} {\bibfnamefont {J.-W.}\
  \bibnamefont {Kim}}, \bibinfo {author} {\bibfnamefont {P.~J.}\ \bibnamefont
  {Ryan}}, \ and\ \bibinfo {author} {\bibfnamefont {J.}~\bibnamefont
  {Chakhalian}},\ }\href {\doibase 10.1063/1.5045756} {\bibfield  {journal}
  {\bibinfo  {journal} {Applied Physics Letters}\ }\textbf {\bibinfo {volume}
  {113}},\ \bibinfo {pages} {081602} (\bibinfo {year}
  {2018}{\natexlab{a}})}\BibitemShut {NoStop}%
\bibitem [{\citenamefont {Middey}\ \emph
  {et~al.}(2018{\natexlab{b}})\citenamefont {Middey}, \citenamefont {Meyers},
  \citenamefont {Kareev}, \citenamefont {Cao}, \citenamefont {Liu},
  \citenamefont {Shafer}, \citenamefont {Freeland}, \citenamefont {Kim},
  \citenamefont {Ryan},\ and\ \citenamefont {Chakhalian}}]{Middey:2018p156801}%
  \BibitemOpen
  \bibfield  {author} {\bibinfo {author} {\bibfnamefont {S.}~\bibnamefont
  {Middey}}, \bibinfo {author} {\bibfnamefont {D.}~\bibnamefont {Meyers}},
  \bibinfo {author} {\bibfnamefont {M.}~\bibnamefont {Kareev}}, \bibinfo
  {author} {\bibfnamefont {Y.}~\bibnamefont {Cao}}, \bibinfo {author}
  {\bibfnamefont {X.}~\bibnamefont {Liu}}, \bibinfo {author} {\bibfnamefont
  {P.}~\bibnamefont {Shafer}}, \bibinfo {author} {\bibfnamefont {J.~W.}\
  \bibnamefont {Freeland}}, \bibinfo {author} {\bibfnamefont {J.-W.}\
  \bibnamefont {Kim}}, \bibinfo {author} {\bibfnamefont {P.~J.}\ \bibnamefont
  {Ryan}}, \ and\ \bibinfo {author} {\bibfnamefont {J.}~\bibnamefont
  {Chakhalian}},\ }\href {\doibase 10.1103/PhysRevLett.120.156801} {\bibfield
  {journal} {\bibinfo  {journal} {Phys. Rev. Lett.}\ }\textbf {\bibinfo
  {volume} {120}},\ \bibinfo {pages} {156801} (\bibinfo {year}
  {2018}{\natexlab{b}})}\BibitemShut {NoStop}%
\bibitem [{\citenamefont {Middey}\ \emph
  {et~al.}(2018{\natexlab{c}})\citenamefont {Middey}, \citenamefont {Meyers},
  \citenamefont {Kareev}, \citenamefont {Liu}, \citenamefont {Cao},
  \citenamefont {Freeland},\ and\ \citenamefont
  {Chakhalian}}]{Middey:2018ENOLNOStrain}%
  \BibitemOpen
  \bibfield  {author} {\bibinfo {author} {\bibfnamefont {S.}~\bibnamefont
  {Middey}}, \bibinfo {author} {\bibfnamefont {D.}~\bibnamefont {Meyers}},
  \bibinfo {author} {\bibfnamefont {M.}~\bibnamefont {Kareev}}, \bibinfo
  {author} {\bibfnamefont {X.}~\bibnamefont {Liu}}, \bibinfo {author}
  {\bibfnamefont {Y.}~\bibnamefont {Cao}}, \bibinfo {author} {\bibfnamefont
  {J.~W.}\ \bibnamefont {Freeland}}, \ and\ \bibinfo {author} {\bibfnamefont
  {J.}~\bibnamefont {Chakhalian}},\ }\href@noop {} {\bibfield  {journal}
  {\bibinfo  {journal} {Phys. Rev. B}\ }\textbf {\bibinfo {volume} {98}},\
  \bibinfo {pages} {045115} (\bibinfo {year} {2018}{\natexlab{c}})}\BibitemShut
  {NoStop}%
\bibitem [{\citenamefont {Medarde}\ \emph {et~al.}(1992)\citenamefont
  {Medarde}, \citenamefont {Fontaine}, \citenamefont {Garc\'{\i}a-Mu\~noz},
  \citenamefont {Rodr\'{\i}guez-Carvajal}, \citenamefont {de~Santis},
  \citenamefont {Sacchi}, \citenamefont {Rossi},\ and\ \citenamefont
  {Lacorre}}]{Medarde:1992p14975}%
  \BibitemOpen
  \bibfield  {author} {\bibinfo {author} {\bibfnamefont {M.}~\bibnamefont
  {Medarde}}, \bibinfo {author} {\bibfnamefont {A.}~\bibnamefont {Fontaine}},
  \bibinfo {author} {\bibfnamefont {J.~L.}\ \bibnamefont
  {Garc\'{\i}a-Mu\~noz}}, \bibinfo {author} {\bibfnamefont {J.}~\bibnamefont
  {Rodr\'{\i}guez-Carvajal}}, \bibinfo {author} {\bibfnamefont
  {M.}~\bibnamefont {de~Santis}}, \bibinfo {author} {\bibfnamefont
  {M.}~\bibnamefont {Sacchi}}, \bibinfo {author} {\bibfnamefont
  {G.}~\bibnamefont {Rossi}}, \ and\ \bibinfo {author} {\bibfnamefont
  {P.}~\bibnamefont {Lacorre}},\ }\href {\doibase 10.1103/PhysRevB.46.14975}
  {\bibfield  {journal} {\bibinfo  {journal} {Phys. Rev. B}\ }\textbf {\bibinfo
  {volume} {46}},\ \bibinfo {pages} {14975} (\bibinfo {year}
  {1992})}\BibitemShut {NoStop}%
\bibitem [{\citenamefont {Liu}\ \emph {et~al.}(2010)\citenamefont {Liu},
  \citenamefont {Kareev}, \citenamefont {Gray}, \citenamefont {Kim},
  \citenamefont {Ryan}, \citenamefont {Dabrowski}, \citenamefont {Freeland},\
  and\ \citenamefont {Chakhalian}}]{Liu:2010p233110}%
  \BibitemOpen
  \bibfield  {author} {\bibinfo {author} {\bibfnamefont {J.}~\bibnamefont
  {Liu}}, \bibinfo {author} {\bibfnamefont {M.}~\bibnamefont {Kareev}},
  \bibinfo {author} {\bibfnamefont {B.}~\bibnamefont {Gray}}, \bibinfo {author}
  {\bibfnamefont {J.~W.}\ \bibnamefont {Kim}}, \bibinfo {author} {\bibfnamefont
  {P.}~\bibnamefont {Ryan}}, \bibinfo {author} {\bibfnamefont {B.}~\bibnamefont
  {Dabrowski}}, \bibinfo {author} {\bibfnamefont {J.~W.}\ \bibnamefont
  {Freeland}}, \ and\ \bibinfo {author} {\bibfnamefont {J.}~\bibnamefont
  {Chakhalian}},\ }\href@noop {} {\bibfield  {journal} {\bibinfo  {journal}
  {Applied Physics Letters}\ }\textbf {\bibinfo {volume} {96}},\ \bibinfo
  {pages} {233110} (\bibinfo {year} {2010})}\BibitemShut {NoStop}%
\bibitem [{\citenamefont {Wu}\ \emph {et~al.}(2015)\citenamefont {Wu},
  \citenamefont {Benckiser}, \citenamefont {Audehm}, \citenamefont {Goering},
  \citenamefont {Wochner}, \citenamefont {Christiani}, \citenamefont
  {Logvenov}, \citenamefont {Habermeier},\ and\ \citenamefont
  {Keimer}}]{Wu:2015p195130}%
  \BibitemOpen
  \bibfield  {author} {\bibinfo {author} {\bibfnamefont {M.}~\bibnamefont
  {Wu}}, \bibinfo {author} {\bibfnamefont {E.}~\bibnamefont {Benckiser}},
  \bibinfo {author} {\bibfnamefont {P.}~\bibnamefont {Audehm}}, \bibinfo
  {author} {\bibfnamefont {E.}~\bibnamefont {Goering}}, \bibinfo {author}
  {\bibfnamefont {P.}~\bibnamefont {Wochner}}, \bibinfo {author} {\bibfnamefont
  {G.}~\bibnamefont {Christiani}}, \bibinfo {author} {\bibfnamefont
  {G.}~\bibnamefont {Logvenov}}, \bibinfo {author} {\bibfnamefont {H.-U.}\
  \bibnamefont {Habermeier}}, \ and\ \bibinfo {author} {\bibfnamefont
  {B.}~\bibnamefont {Keimer}},\ }\href {\doibase 10.1103/PhysRevB.91.195130}
  {\bibfield  {journal} {\bibinfo  {journal} {Phys. Rev. B}\ }\textbf {\bibinfo
  {volume} {91}},\ \bibinfo {pages} {195130} (\bibinfo {year}
  {2015})}\BibitemShut {NoStop}%
\bibitem [{\citenamefont {Middey}\ \emph {et~al.}(2014)\citenamefont {Middey},
  \citenamefont {Rivero}, \citenamefont {Meyers}, \citenamefont {Kareev},
  \citenamefont {Liu}, \citenamefont {Cao}, \citenamefont {Freeland},
  \citenamefont {Barraza-Lopez},\ and\ \citenamefont
  {Chakhalian}}]{Middey:2014p6819}%
  \BibitemOpen
  \bibfield  {author} {\bibinfo {author} {\bibfnamefont {S.}~\bibnamefont
  {Middey}}, \bibinfo {author} {\bibfnamefont {P.}~\bibnamefont {Rivero}},
  \bibinfo {author} {\bibfnamefont {D.}~\bibnamefont {Meyers}}, \bibinfo
  {author} {\bibfnamefont {M.}~\bibnamefont {Kareev}}, \bibinfo {author}
  {\bibfnamefont {X.}~\bibnamefont {Liu}}, \bibinfo {author} {\bibfnamefont
  {Y.}~\bibnamefont {Cao}}, \bibinfo {author} {\bibfnamefont {J.~W.}\
  \bibnamefont {Freeland}}, \bibinfo {author} {\bibfnamefont {S.}~\bibnamefont
  {Barraza-Lopez}}, \ and\ \bibinfo {author} {\bibfnamefont {J.}~\bibnamefont
  {Chakhalian}},\ }\href {http://dx.doi.org/10.1038/srep06819} {\bibfield
  {journal} {\bibinfo  {journal} {Sci. Rep.}\ }\textbf {\bibinfo {volume}
  {4}},\ \bibinfo {pages} {6819} (\bibinfo {year} {2014})}\BibitemShut
  {NoStop}%
\bibitem [{\citenamefont {Freeland}, \citenamefont {van Veenendaal},\ and\
  \citenamefont {Chakhalian}(2016)}]{Freeland:2016p56}%
  \BibitemOpen
  \bibfield  {author} {\bibinfo {author} {\bibfnamefont {J.~W.}\ \bibnamefont
  {Freeland}}, \bibinfo {author} {\bibfnamefont {M.}~\bibnamefont {van
  Veenendaal}}, \ and\ \bibinfo {author} {\bibfnamefont {J.}~\bibnamefont
  {Chakhalian}},\ }\href {\doibase
  https://doi.org/10.1016/j.elspec.2015.07.006} {\bibfield  {journal} {\bibinfo
   {journal} {Journal of Electron Spectroscopy and Related Phenomena}\ }\textbf
  {\bibinfo {volume} {208}},\ \bibinfo {pages} {56 } (\bibinfo {year}
  {2016})},\ \bibinfo {note} {special Issue: Electronic structure and function
  from state-of-the-art spectroscopy and theory}\BibitemShut {NoStop}%
\bibitem [{\citenamefont {Meyers}\ \emph {et~al.}(2016)\citenamefont {Meyers},
  \citenamefont {Liu}, \citenamefont {Freeland}, \citenamefont {Middey},
  \citenamefont {Kareev}, \citenamefont {Kwon}, \citenamefont {Zuo},
  \citenamefont {Chuang}, \citenamefont {Kim}, \citenamefont {Ryan},\ and\
  \citenamefont {et~al.}}]{Meyers:2016p27934}%
  \BibitemOpen
  \bibfield  {author} {\bibinfo {author} {\bibfnamefont {D.}~\bibnamefont
  {Meyers}}, \bibinfo {author} {\bibfnamefont {J.}~\bibnamefont {Liu}},
  \bibinfo {author} {\bibfnamefont {J.~W.}\ \bibnamefont {Freeland}}, \bibinfo
  {author} {\bibfnamefont {S.}~\bibnamefont {Middey}}, \bibinfo {author}
  {\bibfnamefont {M.}~\bibnamefont {Kareev}}, \bibinfo {author} {\bibfnamefont
  {J.}~\bibnamefont {Kwon}}, \bibinfo {author} {\bibfnamefont {J.~M.}\
  \bibnamefont {Zuo}}, \bibinfo {author} {\bibfnamefont {Y.-D.}\ \bibnamefont
  {Chuang}}, \bibinfo {author} {\bibfnamefont {J.~W.}\ \bibnamefont {Kim}},
  \bibinfo {author} {\bibfnamefont {P.~J.}\ \bibnamefont {Ryan}}, \ and\
  \bibinfo {author} {\bibnamefont {et~al.}},\ }\href {\doibase
  10.1038/srep27934} {\bibfield  {journal} {\bibinfo  {journal} {Scientific
  Reports}\ }\textbf {\bibinfo {volume} {6}},\ \bibinfo {pages} {27934}
  (\bibinfo {year} {2016})}\BibitemShut {NoStop}%
\bibitem [{\citenamefont {Post}\ \emph {et~al.}(2018)\citenamefont {Post},
  \citenamefont {McLeod}, \citenamefont {Hepting}, \citenamefont {Bluschke},
  \citenamefont {Wang}, \citenamefont {Cristiani}, \citenamefont {Logvenov},
  \citenamefont {Charnukha}, \citenamefont {Ni}, \citenamefont {Radhakrishnan}
  \emph {et~al.}}]{post:2018p1056}%
  \BibitemOpen
  \bibfield  {author} {\bibinfo {author} {\bibfnamefont {K.}~\bibnamefont
  {Post}}, \bibinfo {author} {\bibfnamefont {A.}~\bibnamefont {McLeod}},
  \bibinfo {author} {\bibfnamefont {M.}~\bibnamefont {Hepting}}, \bibinfo
  {author} {\bibfnamefont {M.}~\bibnamefont {Bluschke}}, \bibinfo {author}
  {\bibfnamefont {Y.}~\bibnamefont {Wang}}, \bibinfo {author} {\bibfnamefont
  {G.}~\bibnamefont {Cristiani}}, \bibinfo {author} {\bibfnamefont
  {G.}~\bibnamefont {Logvenov}}, \bibinfo {author} {\bibfnamefont
  {A.}~\bibnamefont {Charnukha}}, \bibinfo {author} {\bibfnamefont
  {G.}~\bibnamefont {Ni}}, \bibinfo {author} {\bibfnamefont {P.}~\bibnamefont
  {Radhakrishnan}},  \emph {et~al.},\ }\href@noop {} {\bibfield  {journal}
  {\bibinfo  {journal} {Nature Physics}\ }\textbf {\bibinfo {volume} {14}},\
  \bibinfo {pages} {1056} (\bibinfo {year} {2018})}\BibitemShut {NoStop}%
\bibitem [{\citenamefont {Upton}\ \emph {et~al.}(2015)\citenamefont {Upton},
  \citenamefont {Choi}, \citenamefont {Park}, \citenamefont {Liu},
  \citenamefont {Meyers}, \citenamefont {Chakhalian}, \citenamefont {Middey},
  \citenamefont {Kim},\ and\ \citenamefont {Ryan}}]{Upton:2015p036401}%
  \BibitemOpen
  \bibfield  {author} {\bibinfo {author} {\bibfnamefont {M.~H.}\ \bibnamefont
  {Upton}}, \bibinfo {author} {\bibfnamefont {Y.}~\bibnamefont {Choi}},
  \bibinfo {author} {\bibfnamefont {H.}~\bibnamefont {Park}}, \bibinfo {author}
  {\bibfnamefont {J.}~\bibnamefont {Liu}}, \bibinfo {author} {\bibfnamefont
  {D.}~\bibnamefont {Meyers}}, \bibinfo {author} {\bibfnamefont
  {J.}~\bibnamefont {Chakhalian}}, \bibinfo {author} {\bibfnamefont
  {S.}~\bibnamefont {Middey}}, \bibinfo {author} {\bibfnamefont {J.-W.}\
  \bibnamefont {Kim}}, \ and\ \bibinfo {author} {\bibfnamefont {P.~J.}\
  \bibnamefont {Ryan}},\ }\href {\doibase 10.1103/PhysRevLett.115.036401}
  {\bibfield  {journal} {\bibinfo  {journal} {Phys. Rev. Lett.}\ }\textbf
  {\bibinfo {volume} {115}},\ \bibinfo {pages} {036401} (\bibinfo {year}
  {2015})}\BibitemShut {NoStop}%
\bibitem [{\citenamefont {Daptary}\ \emph {et~al.}(2019)\citenamefont
  {Daptary}, \citenamefont {Kumar}, \citenamefont {Kareev}, \citenamefont
  {Chakhalian}, \citenamefont {Bid},\ and\ \citenamefont
  {Middey}}]{Daptary:2019p125105}%
  \BibitemOpen
  \bibfield  {author} {\bibinfo {author} {\bibfnamefont {G.~N.}\ \bibnamefont
  {Daptary}}, \bibinfo {author} {\bibfnamefont {S.}~\bibnamefont {Kumar}},
  \bibinfo {author} {\bibfnamefont {M.}~\bibnamefont {Kareev}}, \bibinfo
  {author} {\bibfnamefont {J.}~\bibnamefont {Chakhalian}}, \bibinfo {author}
  {\bibfnamefont {A.}~\bibnamefont {Bid}}, \ and\ \bibinfo {author}
  {\bibfnamefont {S.}~\bibnamefont {Middey}},\ }\href {\doibase
  10.1103/PhysRevB.100.125105} {\bibfield  {journal} {\bibinfo  {journal}
  {Phys. Rev. B}\ }\textbf {\bibinfo {volume} {100}},\ \bibinfo {pages}
  {125105} (\bibinfo {year} {2019})}\BibitemShut {NoStop}%
\bibitem [{\citenamefont {Park}\ \emph {et~al.}(2013)\citenamefont {Park},
  \citenamefont {Coy}, \citenamefont {Kasirga}, \citenamefont {Huang},
  \citenamefont {Fei}, \citenamefont {Hunter},\ and\ \citenamefont
  {Cobden}}]{Park:2013p431}%
  \BibitemOpen
  \bibfield  {author} {\bibinfo {author} {\bibfnamefont {J.~H.}\ \bibnamefont
  {Park}}, \bibinfo {author} {\bibfnamefont {J.~M.}\ \bibnamefont {Coy}},
  \bibinfo {author} {\bibfnamefont {T.~S.}\ \bibnamefont {Kasirga}}, \bibinfo
  {author} {\bibfnamefont {C.}~\bibnamefont {Huang}}, \bibinfo {author}
  {\bibfnamefont {Z.}~\bibnamefont {Fei}}, \bibinfo {author} {\bibfnamefont
  {S.}~\bibnamefont {Hunter}}, \ and\ \bibinfo {author} {\bibfnamefont {D.~H.}\
  \bibnamefont {Cobden}},\ }\href@noop {} {\bibfield  {journal} {\bibinfo
  {journal} {Nature}\ }\textbf {\bibinfo {volume} {500}},\ \bibinfo {pages}
  {431} (\bibinfo {year} {2013})}\BibitemShut {NoStop}%
\end{thebibliography}

%

 \end{document}